\documentclass[showpacs,amsmath,amssymb,showkeys,pre]{revtex4}
\usepackage{amsmath}
\usepackage{amssymb}
\usepackage{graphics,color,epsfig}

\newcommand{\ud}{\,\mathrm{d}}
\newcommand{\e}{\mathrm{e}}

\newcommand{\bx}{\mathbf{x}}
\newcommand{\by}{\mathbf{y}}
\newcommand{\C}{\mathcal{C}}
\renewcommand{\P}{\mathbb{P}}
\newcommand{\N}{\mathcal{N}}
\newcommand{\E}{\mathbb{E}}
\newcommand{\s}{\sigma}
\renewcommand{\t}{\tau}
\renewcommand{\L}{\mathcal{L}}

\newcommand{\Z}{\mathbb{Z}}
\newcommand{\vers}{\rightarrow}

\newcommand{\btt}{ \mbox{\boldmath$\tau$}}
\newcommand{\bs}{ \mbox{\boldmath$\sigma$}}

\begin{document}

\title{Statistical mechanics of error exponents for error-correcting codes}

\author{Thierry Mora}
\affiliation{Laboratoire de Physique Th\'eorique et Mod\`eles
Statistiques, B\^at.~100, Universit\'e Paris-Sud, F--91405 Orsay,
France.}

\author{Olivier Rivoire}
\affiliation{Laboratory of Living Matter, The Rockefeller
University, 1230 York Avenue, Box 34, New York, NY--10021, USA.}

\begin{abstract}
Error exponents characterize the exponential decay, when increasing message length, of the probability of error of many error-correcting codes. To tackle the long standing problem of computing them exactly, we introduce a general, thermodynamic, formalism that we illustrate with maximum-likelihood decoding of low-density parity-check (LDPC) codes on the binary erasure channel (BEC) and the binary symmetric channel (BSC). In this formalism, we apply the cavity method for large deviations to derive expressions for both the average and typical error exponents, which differ by the procedure used to select the codes from specified ensembles. When decreasing the noise intensity, we find that two phase transitions take place, at two different levels: a glass to ferromagnetic transition in the space of codewords, and a paramagnetic to glass transition in the space of codes.
\end{abstract}


\maketitle

\section{Introduction}

Communicating information requires a physical channel whose inherent noise impairs the transmitted signals. Reliability can be improved by adding redundancy to the messages, thus allowing the receiver to correct the effects of the noise. This procedure has the drawbacks of increasing the cost of generating and sending the messages, and of decreasing the speed of transmission. At first sight, better accuracy seems achievable only at the expense of lesser efficiency. Remarkably, Shannon showed that, in the limit of infinite-length messages, error-free communication is possible using only limited redundancy~\cite{Shannon48}. His proof of principle has triggered many efforts to construct actual error-correcting schemes that would approach the theoretical bounds.  A renewal of interest for the subject has taken place during the last ten years, as new error-correcting codes were finally discovered~\cite{BerrouGlavieux93}, or rediscovered~\cite{MacKay03}, which showed practical performances close to Shannon's bounds.

In this paper, we analyze a major family of such codes, the low-density parity-check (LDPC) codes, also known as Gallager codes, from the name of their inventor~\cite{Gallager62}.  Our focus is on the characterization of rare decoding errors, in situations where most realizations of the noise are accurately corrected. Error-free communication, as guaranteed by Shannon's theorem, indeed results from a law of large number, and is achieved only with infinite-length messages. Accordingly, any error-correcting scheme acting on finite-length messages has a non-zero error probability, which generically vanishes exponentially with the message length. Such error probabilities are described by {\it error exponents}, giving their rate of exponential decay.  Two kinds of error exponents are usually distinguished: {\it average} error exponents, where the average is taken over an ensemble of codes, and {\it typical} error exponents, where the codes are typical elements of their ensemble.

The study of error exponents has attracted early on considerable attention in the information theory community, but exact expressions have turned particularly difficult to derive (see e.g.~\cite{Verdu98} and~\cite{Berlekamp02} for concise and non-technical reviews with entries in the literature). Exact asymptotic results are known in the limit of the so-called {\it random linear model}~\cite{BargForney02} (presented in Appendix \ref{sec:RLM}), but only loose bounds (presented in Appendix \ref{sec:union}) have been established for more general codes. Recently, a systematic finite-length analysis of LDPC codes under iterative decoding was carried out for the binary erasure channel (BEC)~\cite{DiProietti02,AmraouiMontanari04}, yielding exact, yet non-explicit, formul{\ae} for the average error probability. Up to now, little has however been known of the error probability under maximum-likelihood decoding, except for the work of~\cite{SkantzosvanMourik03} dealing with the binary symmetric channel (BSC). 

We address here the problem of computing error exponents of LDPC codes under maximum-likelihood decoding, over both the BEC and BSC (all the necessary definitions are recalled below). We adopt a statistical physics point of view, which exploits the well established~\cite{NishimoriBook01} mapping between error-correcting codes and spin glasses~\cite{Sourlas89}. A thermodynamic formalism is introduced where error exponents are expressed as large deviation functions~\cite{denHollander00}, which we compute by means of the extension of the cavity method~\cite{MezardParisi01} proposed in~\cite{Rivoire05}. This approach offers an alternative to the related replica method employed in~\cite{SkantzosvanMourik03} and allows us to address both average and typical error exponents.  We thus obtain an interesting phase diagram, with two very distinct phase transitions occurring when the intensity of the noise in the channels is varied.

A brief summary of our results can be found in~\cite{MoraRivoire06a}. We present in what follows a much more detailed account of our approach. In a first part, we define LDPC codes, recall their mapping to some models of spin glasses and optimization problems, and give a general overview of our thermodynamic (large deviation) formalism. The two subsequent parts apply this framework to the analysis of LDPC codes over the binary erasure channel (BEC) and the binary symmetric channel (BSC) respectively. We sum up our results in a conclusion where we also point out some open questions. Most of the technical calculations are relegated to the appendices, which also contain a detailed discussion of the limiting case of random linear codes.

\section{Error correcting codes and the large deviation formalism}\label{sec:part2}

\subsection{Error correcting codes}

\begin{figure}
\begin{center}
\epsfig{file=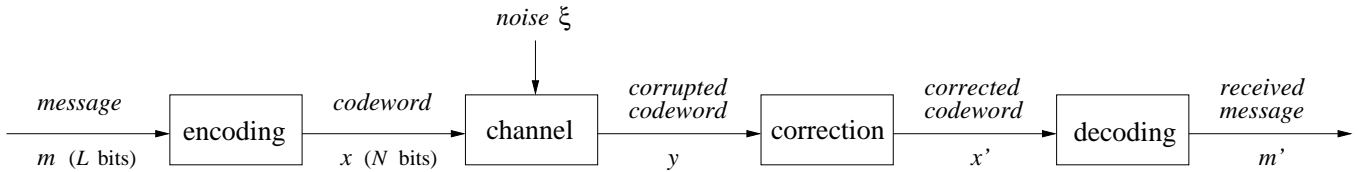,width=\linewidth}
\caption{\label{fig:scheme}\small Error correction scheme. A message $m$ composed of $L$ bits, $m\in\{0,1\}^L$, is first encoded in a codeword of longer size $N$ with $R=L/N>1$, defining the rate of the code. The noise $\xi$ of the channel corrupts the transmitted codeword which becomes $y$ (see Fig.~\ref{fig:channel} for examples of channels). This output is generically not a codeword, and the correction consists in inferring the most probable codeword to which it comes from. Finally, the inferred codeword $x'$ is converted back into its corresponding message $m'$. The communication is successful if $m'=m$.}
\end{center}
\end{figure}

Error correcting codes are based on the idea that adding sufficient
redundancy to the messages can allow the receiver to reconstruct them,
even if they have been partially corrupted by the
noisy channel~\cite{CoverThomas91}. A schematic view of how these
codes operate is presented in figure~\ref{fig:scheme}.
Given a {\it message} composed of $L$ bits, an encoding map
$\{0,1\}^L\to\{0,1\}^N$ first introduces redundancy by converting
the $L$ bits of the message into a longer sequence of $N$ bits,
called a {\it codeword}. The ratio $R\equiv L/N$ defines the {\it
rate} of the code, and should ideally be as large as possible to
reduce communication costs, yet small enough to allow for
corrections. Corrections are implemented downstream the noisy
channel and specified by a decoding map $\{0,1\}^N\to\{0,1\}^L$
whose purpose is to reconstruct the original message from the
received corrupted codeword. Decoding is composed of two steps:
first, the most probable codeword is inferred, and second, it is
converted into its corresponding message.

In this scheme, messages and codewords are related by the
one-to-one encoding map, and translating messages into codewords
or conversely is relatively straightforward. The
computationally most demanding part is concentrated on inferring the
most probable codeword sent, given the corrupted
codeword received. In what follows, we shall focus
exclusively on this problem, which requires manipulating only
codewords.

\subsection{Communication channels}

Formally, a noisy channel is characterized by a transition
probability $Q(\by|\bx)$ giving the probability for its output to
be $\by$ given that its input was $\bx$. For the sake of simplicity,
we confine to {\it memoryless} channels where the noise
affects each bit independently of the others, i.e.,
$Q(\by|\bx)=\prod_{i=1}^NQ(y_i|x_i)$ with $Q(y_i|x_i)$ independent
of $i$.

\begin{figure}
\begin{minipage}[b]{.46\linewidth}
\begin{center}
\epsfig{file=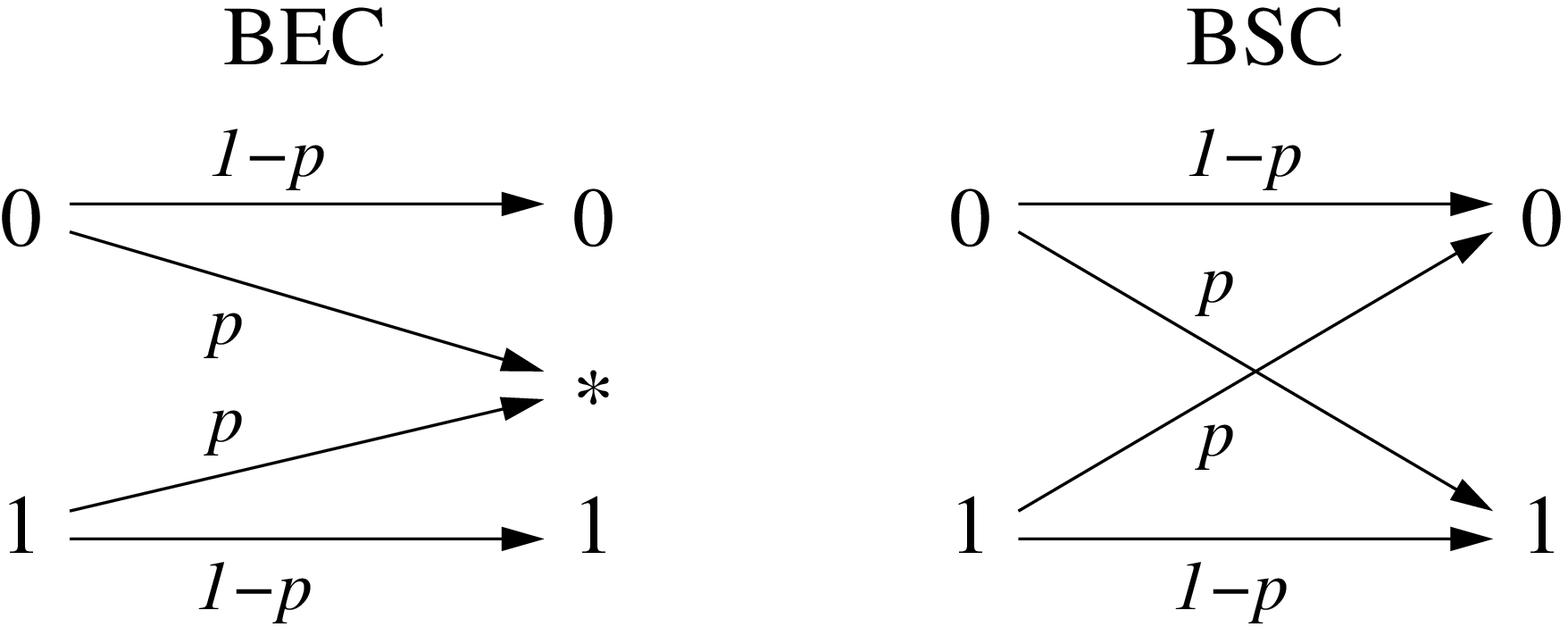,height=3cm}
\caption{\label{fig:channel}\small Communication channels. On the left the BEC (binary erasure channel) erases a bit with probability $p$ and leaves it unchanged with probability $1-p$. On the right the BSC (binary symmetric channel) flips a bit with probability $p$ and leaves it unchanged with probability $1-p$.}
\end{center}
\end{minipage}\hfill
\begin{minipage}[b]{.46\linewidth}
\begin{center}
\epsfig{file=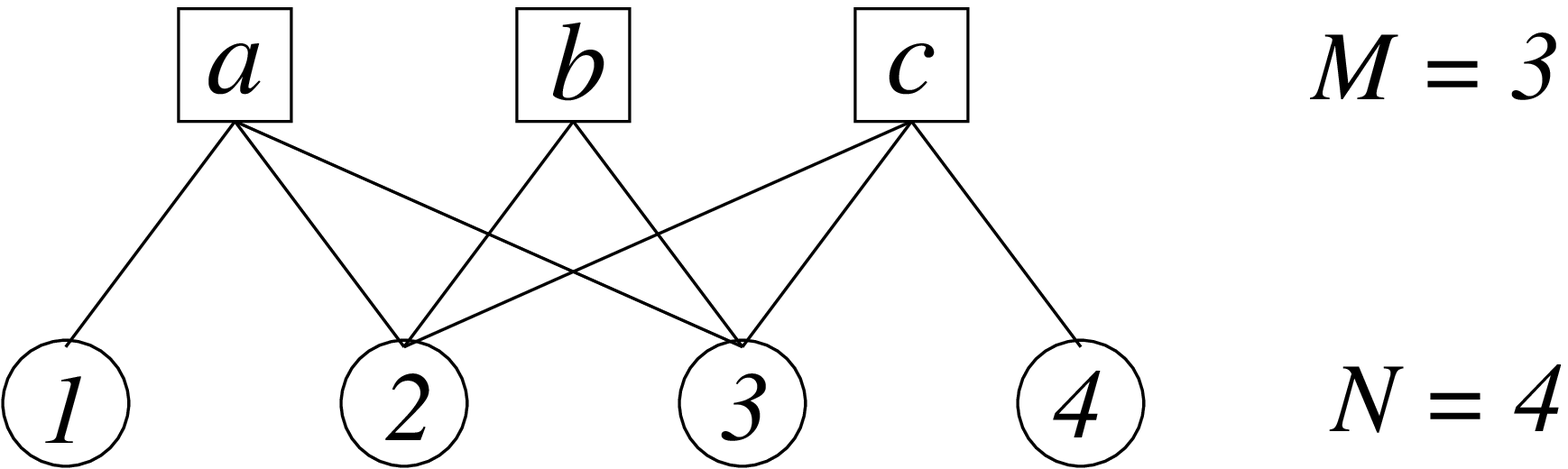,width=.9\linewidth}
\caption{\label{fig:factorgraph}\small Factor graph (Tanner graph \cite{Tanner81}). The circles represent the variable nodes, associated with the $N$ bits $\{x_i\}$, and the square represent the $M$ parity-check. In the example given, the constraints read: $(a)$: $x_1+x_2+x_3=0$, $(b)$: $x_2+x_3=0$, $(c)$: $x_2+x_3+x_4=0$ (modulo 2).}
\end{center}
\end{minipage}
\end{figure}

We shall consider more specifically two examples of memoryless
channels. The first one is the {\it binary erasure channel} (BEC)
where a bit is erased with probability $p$, that is $Q(*|x)=p$ and
$Q(x|x)=1-p$ where $*$ represents an erased bit (see
Fig.~\ref{fig:channel}). The second is the {\it binary symmetric
channel} (BSC) where a bit is flipped with probability $p$, that
is $Q(0|1)=Q(1|0)=p$ and $Q(0|0)=Q(1|1)=1-p$ (see
Fig.~\ref{fig:channel}).

\subsection{LDPC codes and code ensembles}

Shannon first formalized the problem of error correction and
determined the lowest achievable rate $R$ allowing error-free
correction~\cite{Shannon48}. He found a general expression for this limit, called the
{\it channel capacity}, which depends only on the nature of the channel and takes the form $C_{\rm
BEC}(p)=1-p$ and $C_{\rm BSC}(p)=1-p\ln p-(1-p)\ln(1-p)$ for the
BEC and BSC respectively. Shannon's proof for the
existence of codes achieving the channel capacity was
non-constructive, and his analysis restricted to the
limit of infinitely long messages, $L\to\infty$. Amongst the various families of codes proposed to practically perform error correction, one of the most promising is the family of {\it low-density parity-check} (LDPC) codes~\cite{Gallager62}.

A LDPC code is defined by a sparse matrix $A$ where ``sparse'' means
that $A$ is mostly composed of 0's, with otherwise a few 1's.
The {\it parity-check matrix} $A$ has size $M\times N$ with
$M=N-L$, and is associated with a {\it generator matrix} $G$ of
size $L\times N$ such that $GA=0$ (see e.g. \cite{MacKay03} for
explicit constructions); the encoding map is taken to be the
linear map $x=Gm$ and the rate of the code is $R=L/N=1-M/N$. By
construction, a $N$-bit codeword $x$ satisfies the $M$
parity-check equations $Ax=0$, or, in other words, the set of
codewords is the kernel of $A$. The parity-check matrix $A$ is
usually represented graphically by a {\it factor graph}, as in
figure~\ref{fig:factorgraph}: the columns of $A$ are associated
with {\it check nodes} labeled with $a\in\{1,\dots,M\}$, and
represented by squares, and the lines of $A$ are associated with
{\it variable nodes} labeled with $i\in\{1,..,\dots N\}$, and
represented by circles. A non-zero element of the matrix $A$ such
as $A_{ia}=1$ appears as a link between the variable node $i$ and
the check node $a$.

A particularly powerful approach for analyzing error-correcting
codes is the probabilistic method where, instead of considering
a single code, one studies an {\it ensemble} of codes. 
With LDPC codes, code ensembles corresponds to
sets of matrices, or, equivalently, sets of factor graphs. A popular
choice is to consider the ensemble of factor graphs with given
connectivities $c_k$ and $v_\ell$, that is the set of factor
graphs having $c_kM$ check nodes with connectivity $k$ and
$v_\ell N$ variable nodes with connectivity $\ell$, where
$\sum_kc_k=\sum_\ell v_\ell=1$. A convenient representation is by means of
the generating functions $c(x)=\sum_k c_kx^k$ and $v_\ell=\sum_\ell
v_\ell x^\ell$; these notations
allow for instance to write the mean connectivities as $\langle
k\rangle=c'(1)$ and $\langle\ell\rangle=v'(1)$. Due to their
simplicity, a particular attention will be devoted to {\it
regular} codes, whose check nodes have all same degree $k$ and
variable nodes same degree $\ell$, corresponding to
$c_{k'}=\delta_{k,k'}$ and $v_{\ell'}=\delta_{\ell,\ell'}$, or,
equivalently, $c(x)=x^k$ and $v(x)=x^\ell$.

The mathematical fact underlying the probabilistic method is the
phenomenon of {\it measure concentration} which occurs in the
limit where $N\to\infty$ and $M\to\infty$ with fixed ratio
$\alpha=M/N$: in this limit, many properties are shared by {\it almost all}
elements of the ensemble (i.e., all but a subset of measure zero). 
As a consequence, by studying average properties over an
ensemble, one actually has access to properties of typical
elements of this ensemble. This fact is one of the building blocks
of random graph theory~\cite{Bollobas01} and is also central to the physics of
disordered systems where it is known as the {\it
self-averaging property}~\cite{MezardParisi87b}.

While the factor graph representation makes obvious the connection
between LDPC codes and random graph theory, it will also turn particularly
fruitful to exploit the close ties of LDPC codes with both
optimization problems~\cite{PapadimitriouSteiglitz82} and spin-glass
systems~\cite{MezardParisi87b}. LDPC codes are indeed intimately
related to a class of combinatorial optimization problems known as XORSAT problems where, given a sparse matrix $A$ and a
vector $\tau$, one is to find solutions $\s$ to the equation $A\s=\tau$. Although algorithmically relatively simple (Gauss method provides an answer in a time
polynomial in the size of the matrix), XORSAT problems share many
common features with notably more difficult, NP-complete~\cite{PapadimitriouSteiglitz82}, problems such as $K$-SAT. A recent physical approach to XORSAT problems makes use of their formal equivalence with a class of spin-glass systems known as
$p$-spin models~\cite{RicciWeigt01,CoccoDubois03,MezardRicci03}. We
shall follow this line of investigation and apply the {\it cavity
method}~\cite{MezardParisi01,MezardParisi03} from spin-glass theory to
analyze LDPC codes.  We note that alternative, sometimes equivalent,
physical approaches have previously been applied to LDPC codes; we
refer the reader to~\cite{KabashimaSaad04} for a review of the subject.

The distinctive feature of XORSAT at the root of its
computational simplicity is the presence of an underlying group
symmetry that relates all solutions. In the context of LDPC codes,
it corresponds to the fact that the set of codewords is the
kernel of the parity-check matrix $A$; we shall refer to the XORSAT problem $A\s=0$ whose solutions define the set of codewords as the {\it encoding-CSP} of the LDPC code with check matrix $A$ (CSP stands for {\it constraint satisfaction problem}). The group symmetry has a number of interesting consequences which will
crucially simplify the analysis.

Most of the interest for LDPC codes stems from the possibility to decode them using efficient, iterative algorithms (described in Sec.~\ref{sec:BP}). Unless otherwise stated, we shall however be here concerned with the theoretically simpler, yet computationally much more demanding, {\it maximum-likelihood decoding} procedure. It consists in systematically decoding a received message to the most probable codeword (a task that iterative algorithms are in some cases unable to perform, as recalled in Sec.~\ref{sec:BP}).

Finally, it is interesting to note that in the limit where $\langle
k\rangle,\langle\ell\rangle\to\infty$ with fixed ratio, LDPC codes
define the {\it random linear model} (RLM) whose typical elements have
been shown by Shannon to achieve the channel capacity. This
particular limit, where many quantities can be computed by invoking
only elementary combinatorial arguments, is discussed in details in
appendix~\ref{sec:RLM}.

\subsection{Typical properties and phase transitions}

The performance of a particular code over a given channel is
measured by its error probability i.e., the probability that it
fails to correctly decode a corrupted codeword. More precisely, if
$d(\by)$ denotes the inferred codeword when $\bx$ is sent and $\by$ received, one defines the
{\it block error probability} for $\bx$ as
\begin{equation}
\P_N^{(B)}(\bx)=\sum_\by Q(\by|\bx)1_{d(\by)\neq \bx},
\end{equation}
and the average block error probability as
\begin{equation}
P_N^{(B)}=\E_{\bx}[\P_N^{(B)}(\bx)],
\end{equation}
where $\E_x$ denotes the expectation (average) over the set of
codewords. With LDPC codes, this average is trivial since, due to
the group symmetry, all codewords are equivalent, and
$\P_N^{(B)}(\bx)$ is independent of $\bx$.

The concentration phenomenon alluded above means
here that $P_N^{(B)}\to p_B$ with $N\to\infty$ within a given code
ensemble defined by generating functions $c(x)$ and $v(x)$. As the
level of the noise $p$ is increased, a phase transition is
generically observed: a critical value $p_c$ exists above which
error-free correction is no longer possible ($p_B=0$ for $p<p_c$ and $p_B=1$ for $p>p_c$). The formalism to be
presented in the next sections will yield in particular the value
of $p_c$ for given code ensembles and channels. Obviously, the
presence of this phase transition indicates that, when using a
channel with noise level $p$, one should choose a code from an
ensemble for which $p<p_c$. The phase transition is however
occurring only in the limit of infinite codewords (thermodynamic
limit) whereas practical coding inevitably deals with finite $N$.
This leads to the fact that the block error probability is
not exactly zero, even in the regime $p<p_c$.

For a given code of finite but large block-length $N$, error can thus be caused by rare, atypical, realizations of the noise.
Similarly, when picking a code at random from a code ensemble of
finite size, one can observe properties differing from the typical
properties predicted by the law of large numbers. 
We show in what follows how these two atypical features induced by finite-size
effects can be analyzed in a common framework.

\subsection{Large deviations}\label{sec:overview}

At this stage, it is useful to make explicit the three different
levels of statistics involved in the analysis
of error-correcting codes: $(i)$ Statistics over the codes $\C$ in
a defined code ensemble $C$; $(ii)$ Statistics over the set of
transmitted codewords $\bx$ of a particular code; $(iii)$ Statistics over the noise $\xi$ of the channel, with a specified $p$.
For given $\C$, $\bx$, $\xi$, a fourth level of statistics is involved in the decoding process, over the possible codewords $y\in\{0,1\}^N$ from which the received corrupted codeword originates. The group structure of the
set of codewords of LDPC codes makes the level $(ii)$ trivial
since all codewords are in fact equivalent (isomorphic). We will
consequently ignore it and address only the levels $(i)$ and
$(iii)$.

The problem of evaluating the probability that, due to finite-size
effects, a property differs from the typical case belongs to {\it large deviation} theory~\cite{denHollander00}. To give here a general
presentation of the concepts and methods to be used, we assume
that the success of the decoding is measured by a function
$S_N(\xi,\C)$ extensive in $N$, and such that
$S_N(\xi,\C)\leq 0$ if the code $\C$ correctly decodes a message subject to noise $\xi$,
and $S_N(\xi,\C)>0$ otherwise; in the next sections, we will show
explicitly how such an observable can be defined with LDPC codes, both for the
BEC and the BSC channels. In terms of $S_N$, the decoding phase
transition takes the following form: in the limit $N\to\infty$, the distribution of the density $S_N/N$ concentrates around a typical value $s_{\rm typ}(p)$ which verifies $s_{\rm typ}(p)\leq 0$ if $p<p_c$, and $s_{\rm typ}(p)>0$ if $p>p_c$, where $p$ denotes as before the level of noise of the channel (see Fig.~\ref{fig:channel} for examples).

For typical codes in their ensemble, denoted $\C^0$, we describe large deviations of $S_N$ with respect to the noise $\xi$ by a {\it rate function} $L_0(s)$ such that the probability to observe $S_N(\xi,\C^0)/N=s$ satisfies
\begin{equation}\label{eq:L_C}
\P_N[\xi:S_N(\xi,\C^0)/N=s]\asymp e^{-NL_0(s)}.
\end{equation}
Here the symbol $a_N\asymp b_N$ refers to an exponential equivalence,
$\ln a_N/\ln b_N\to 1$ as $N\to\infty$. Viewed as a function of the noise
level $p$, the rate function $E_{\rm typ}(p)=L_0(s=0)$ is known in the coding
literature as the {\it typical error exponent}~\cite{Verdu98}. The exponential
decay with $N$ of atypical properties is quite generic when dealing
with large deviations, but this scaling is not necessarily insured, as discussed in more details in appendix~\ref{sec:scaling}. In the thermodynamic formalism that we shall adopt, rate functions are computed by introducing a potential $\Phi_{\C}(x)$ defined by 
\begin{equation}\label{eq:phiC}
\Phi_{\C}(x)=\ln\left(\E_\xi[e^{xS_N(\xi,\C)}]\right).
\end{equation}
In the limit $N\to\infty$ limit , the density $\Phi_\C(x)/N$ tends to a typical value $\phi_0(x)$, which is related to the rate function $L_0(s)$ by
\begin{equation}
e^{N\phi_0(x)}\asymp\int ds\ e^{N[xs-L_0(s)]}.
\end{equation}
Equivalently, by taking the saddle point,
\begin{equation}\label{eq:Legendre}
\phi_0(x)=xs-L_0(s),\quad x=\partial_sL_0(s).
\end{equation}
The rate function $L_0(s)$ can thus be reconstructed from $\phi_0(x)$ by inverting the Legendre transformation,
\begin{equation}
L_0(s)=sx-\phi_0(x),\quad s=\partial_x\phi_0(x).
\end{equation}
The analogy with usual thermodynamics is summarized in table~\ref{tab:analogy}.

\begin{table}
\begin{tabular}{l||c|c|c|c|c}
 & \emph{Spin glass} & \emph{Average} & \emph{Typical}
 & \emph{Multi-step, step 1} & \emph{Multi-step, step 2} \\
 \hline\hline
Disorder & couplings $J_{ij}$ & - & typical codes $\C^0$ & codes $\C$ at $y$ & -\\
\hline Configurations & spins $\{\sigma_i\}_i$ & noise+codes $(\xi,\C)$ &
noise $\xi$ & noise $\xi$  & codes $\C$\\
\hline
Observable & $E=\sum_{ij}J_{ij}\sigma_i\sigma_j$ & $S_N(\xi,\C)$ & $S_N(\xi,\C^0)$  & $S_N(\xi,\C)$ & $L_\C(s)$ \\
\hline Entropy & $s(e=E/N)$ & $L_0(s=S_N/N)$  & $L(s=S_N/N)$ & $\mathcal{L}(\phi,x)$\\
\hline Temperature$^{-1}$ & $\beta=\partial_e s$ & $x=\partial_sL_1$ & $x=\partial_s L_0$ & $x=\partial_sL$ & $y=\partial_\phi\mathcal{L}$\\
\hline Potential & $\beta f=\beta e-s$ & $\phi_1=xs-L_1$ &
$\phi_0=xs-L_0$ & $\phi=xs-L$ & $\psi=y\phi-\mathcal{L}$
\end{tabular}
\caption{This table presents the analogy with spin glasses or, more generally, the statistical physics of disordered system with quenched disorder.}\label{tab:analogy}
\end{table}

From a theoretical perspective, it is simpler to make an average over the codes and compute the rate function $L_1(s)$ defined as
\begin{equation}
\P_N[\xi,\C:S_N(\xi,\C)/N=s]\asymp e^{-NL_1(s)}.
\end{equation}
This procedure yields the so-called {\it average error exponent}, $E_{\rm av}=L_1(s=0)$. In the thermodynamical formalism, $L_1(s)$ is conjugated to the potential $\phi_1(x)$ satisfying
\begin{equation}\label{eq:defphi}
e^{N\phi_1(x)}=\E_{(\xi,\C)}[e^{xS_N(\xi,\C)}]=\int ds\
e^{N[xs-L_1(s)]}.
\end{equation}
The two rate functions $L_0(s)$ and $L_1(s)$ may differ, meaning that the average exponent can be associated with atypical codes. Such atypical codes correspond themselves to large deviations of the potential $\Phi_C(x)$. For fixed values of $x$, we define a rate function $\L(\phi,x)$ as
\begin{equation}\label{eq:Lphis}
\P_N[\C:\Phi_\C(x)/N=\phi]\asymp \e^{-N\mathcal{L}(\phi,x)}.
\end{equation}
In a thermodynamic formalism, $\mathcal{L}(\phi,x)$ is again associated with a potential $\psi(x,y)$ defined by
\begin{equation}\label{eq:defpsi}
e^{N\psi(x,y)}=\E_\C\left[\left(\E_\xi[e^{xS_N(\xi,\C)}]\right)^y\right]=\E_\C[e^{y\Phi_\C(x)}]=\int\ud\phi\ e^{N[y\phi-\mathcal{L}(\phi,x)]}.
\end{equation}
We refer to this hierarchical embedding of large deviations as a {\it multi-step large deviation} structure~\cite{Rivoire05}, a term meant to reflect the formal equivalence with
the multi-step replica symmetry breaking scenario developed for spin glasses~\cite{MezardParisi87b} (see table~\ref{tab:replica}). In the limit $N\to\infty$ where the integral is dominated by its saddle point we obtain the Legendre transformation
\begin{equation}
\psi(x,y)=y\phi-\mathcal{L}(\phi,x),\qquad y=\partial_\phi\mathcal{L}(\phi,x).
\end{equation}

Within this extended framework, we recover the average case by taking $y=1$. Indeed, from the definitions (\ref{eq:defphi}) of $\phi_1(x)$ and (\ref{eq:defpsi}) of $\psi(x,y)$ it follows that
\begin{equation}\label{eq:psidef}
e^{N\psi(x,1)}=\E_C[\E_\xi e^{S_N(\xi,\C)}]=\E_{(\xi,\C)}[e^{xS_N(\xi,\C)}]=e^{N\phi_1(x)}
\end{equation}
that is,
\begin{equation}
\psi(x,y=1)=\phi_1(x).
\end{equation}

This average case differs in general from the typical case which corresponds to $y=0$. Indeed, by definition [see Eq.~(\ref{eq:Lphis})], typical codes are associated with the potential $\phi_0$ minimizing $\mathcal{L}(\phi,x)$, with $\mathcal{L}(\phi_0,x)=0$, yielding $y=\partial_\phi\mathcal{L}=0$.
Note that the potential $\phi_0$ is related to $\psi(x,y)$ by $\phi_0(x)=\lim_{y\to 0}(1/y)\psi(x,y)$, which can also be viewed as a corollary of G\"artner-Ellis theorem~\cite{denHollander00}, best known in statistical physics as the replica trick~\cite{MezardParisi87b}  (see table~\ref{tab:replica}). In the language of the replica method, the average case ($y=1$) and the typical case ($y=0$) are respectively referred to as the annealed and quenched computations.

\begin{table}
\begin{tabular}{c|c}
\emph{Replica (symmetric) theory of spin glasses}  & \emph{Multi-step large deviations for LDPC codes}\\
 \hline\hline
Hamiltonian $H_J[\sigma]=\sum_{ij}J_{ij}\sigma_i\sigma_j$ &
$S_N(\xi,\C)$\\
\hline Disorder $\{J_{ij}\}_{ij}$ & Codes $\C$\\
\hline Configurations $\{\sigma_i\}_i$ & Noise $\xi$\\
\hline Number of replicas $n$ & Temperature$^{-1}$ $y$\\
\hline Physical temperature$^{-1}$ $\beta$ & Temperature$^{-1}$ $x$ \\
\hline Annealed approximation $n=1$ & Average codes $y=1$\\
\hline Quenched computation $n\to 0$ & Typical codes $y\to 0$
\end{tabular}
\caption{Analogy with the replica approach of spin glasses. The replica symmetric method prescribes that the typical partition function $Z_0$ of a disordered system is given by $Z_0\sim\E[Z_N^n]^{1/n}$ with $n\to 0$ or, more precisely, if $\Lambda_N=\ln Z_N$, the typical value of $\lambda=\Lambda_N/N$ is $\lambda_0=\lim_{n\to 0}\lim_{N\to\infty}(1/Nn)\ln\E[e^{n\Lambda_N}]$. This is mathematically justified by the G\"ardner-Ellis theorem which moreover provides a rigorous basis for the interpretation of non-zero values of $n$ in terms of large deviations, as discussed in the text. According to this theorem, if the function $\phi(x)=\lim_{N\to\infty}(1/N)\ln\E[e^{x\Lambda_N}]$ exists and is regular enough (see e.g.~\cite{denHollander00} for a rigorous presentation), then a large deviation principle holds for $\lambda$ with a rate function being the Legendre transform of $\phi(x)$; if we assume the functions differentiable, $L(\lambda)=\lambda x-\phi(x)$ with $\lambda=\partial_x\phi(x)$. As a corollary of this theorem, the typical value $\lambda_0$, which by definition satisfies $L(\lambda_0)=0$ and $x=\partial_\lambda L(\lambda_0)=0$, is given by $\lambda_0=\partial_x\phi(x=0)=\lim_{x\to 0}[\phi(x)/x](x=0)$, as predicted by the replica method. Note also that $n=1$, with $Z_1= \E[Z_N]$, corresponds to the so-called annealed approximation.}\label{tab:replica}
\end{table}

The previous discussion assumed that the potentials were analytical functions of their parameters $x$ and $y$, but this may not be the case, and we will find that phase transitions  can occur when these temperatures are varied. In such cases, taking naively the limit $y\to 0$ leads to erroneous results. We will discuss how to overcome such difficulties when encountering them.  


\section{LDPC codes over the BEC}

We now proceed to illustrate our formalism with LDPC codes over the binary erasure channel (BEC). We start with rederiving the typical phase diagram by means of the cavity method, a slightly different approach than the replica method originally used in~\cite{Montanari01}. This sets the stage for the analysis of error exponents that follows.


\subsection{Typical phase diagram}

\subsubsection{Formulation}

Consider a LDPC code $\C$ with parity-check matrix $A$; its {\it encoding-CSP} (the constraint satisfaction problem whose SAT-assignments define the
codewords) has cost function
\begin{equation}
H_\C[\s]=\sum_{a=1}^M E_a[\s],\qquad {\rm with}\quad
E_a[\s]=\sum_{i=1}^N A_{ai}\s_i\quad (\textrm{mod }2).
\end{equation}
Since $E_a[\s]\in\{0,1\}$, the cost function $H_\C[\s]$ counts the
number of constraints violated by the assignment
$\s=\{\s_i\}_{i=1,\dots,N}$ (where $\s_i\in\{0,1\}$). When a
codeword $\s^*$, satisfying $H_\C[\s^*]=0$, goes through a BEC,
each of its bits $\s_i$ has probability $p$ to be erased. A given
realization of the noise can be characterized by a vector
$\xi=(\xi_1,\dots,\xi_N)$ with $\xi_i=1$ implying that the bit
$\s^*_i$ is lost, and $\xi_i=0$ that it is unaffected. If we denote
by $\mathcal{E}$ the set of indices $i$ for which $\xi_i=1$
(erased bits), the decoding task consists in reconstructing
$\{\s_i^*\}_{i\in\mathcal{E}}$ from the received bits
$\{\s_i^*\}_{i\notin\mathcal{E}}$ and the knowledge of the
encoding-CSP $H_\C$. This decoding problem defines a new constraint
satisfaction problem, the {\it decoding-CSP}, obtained from the
encoding-CSP by fixing the values of the non-corrupted bits. More
explicitly, the decoding-CSP has cost function
$H^{(\xi)}_\C[\s^{(\xi)}]=\sum_a E^{(\xi)}_a[\s^ {(\xi)}]$ where
$\s^{(\xi)}=\{\s_i\}_{i\in\mathcal{E}}$ and
\begin{equation}
E^{(\xi)}_a[\s^{(\xi)}]=\sum_{i\in\mathcal{E}}A_{ai}\s_i+\sum_{i\notin\mathcal{E}}A_{ai}\s^*_i\quad (\textrm{mod }2).
\end{equation}
Decoding is possible if and only if $\{\s^*_i\}_{i\in\mathcal{E}}$
is the only SAT-assignment of the decoding-CSP. 

If $\N_N(\xi,\C)$ denotes the number of solutions of the decoding-CSP, $S_N(\xi,\C)$ 
can be taken as $S_N(\xi,\C)\equiv \ln\N_N(\xi,\C)$. This entropy
fulfills the desired properties, namely $S_N(\xi,\C)\leq 0$ if
decoding is successful, and $S_N(\xi,\C)>0$ otherwise. 

The particularity of LDPC codes compared to other error-correcting
 codes is that the decoding-CSP has same form as the
encoding-CSP (both are XORSAT problems). As a consequence, the
$\Z_2$-symmetry  of the group of codewords is always preserved, at
variance with what  happens in other CSPs where fixing variables breaks
a symmetry. The BEC is also particular compared with other
channels, since the set $\mathcal{E}$ of corrupted bits is known to
the receiver (this will not be the case with the BSC, where
identifying the corrupted bits is  part of the decoding problem).
This entails that bits can only be fixed  to their correct value.

\subsubsection{Cavity approach}\label{sec:typBEC}

Before considering large deviations, it is instructive to recall the typical
results, i.e. the values taken by $S_N(\xi,\C^0)$ when $\C^0$ is a typical code from a
given ensemble specified by $c(x)$ and $v(x)$, and $\xi$ a typical
realization of the noise from the probability distribution
specified by $p$. We resort here to the {\it cavity
method at zero temperature}~\cite{MezardParisi01}, whose validity is based on the
tree-like structure of the factor graphs associated with typical
LDPC codes. The essentially equivalent replica method has been used in the past: in~\cite{FranzLeone02}, $S_N(\xi,\C)$ is thus obtained by first computing a free energy
with the replica method, and then taking the zero temperature
limit to obtain $S_N(\xi,\C)$, viewed as the entropy of the
zero-energy ground states. 

The approach we follow here, which corresponds to a particular implementation of the entropic cavity method presented in~\cite{MezardPalassini05}, has several
advantages over the replica approach: it involves neither
a zero-replica limit nor a zero-temperature limit, it emphasizes
the specificities of LDPC codes associated with the underlying $\Z_2$
symmetry, and it naturally connects to the algorithmic analysis of
single codes. In the common language of the replica and cavity methods, the calculation to be done is coined {\it one-step replica symmetry breaking} (1RSB), and the entropy $s=S_N/N$ is referred to as a {\it complexity}. This is reflected in what follows by the fact that we strictly restrict to SAT assignments and assume that all constraints are satisfied (the reweighting parameter $\mu$, as denoted in~\cite{MezardParisi03}, is here infinite, $\mu=\infty$). This 1RSB approach is known to exactly describe XORSAT problems~\cite{MezardRicci03,CoccoDubois03}.

Let  $P_i(\sigma_i)$ be the probability, taken over
the set of solutions of  the decoding-CSP, that the bit $i$
assumes the value  $\sigma_i\in\{0,1\}$. Due to the preservation
of the $\Z_2$-symmetry,
 no bit can be non-trivially biased: either it is fixed to 0 or
$1$, corresponding to $P_i=\delta_0$ and $P_i=\delta_{1}$
respectively, or it is  completely free, corresponding to
$P_i=(\delta_0+\delta_{1})/2$, where we denote
$\delta_\tau(\sigma)=\delta_{\tau,\sigma}$. In technical terms,
the  evanescent fields that are generically required to compute
entropies  in CSP~\cite{MezardPalassini05} have here a trivial
distribution, thus  explaining that they can be safely ignored, as was
done in~\cite{FranzLeone02}.    

Let $\nu$ be the probability, taken over the $N$
nodes of a typical factor graph, that a bit $i$ is free, i.e. that
$P_i=(\delta_0+\delta_{1})/2$.  Since a free node has equal
probability to be $0$ or $1$, its  contribution to the entropy is
$\ln 2$, and the mean entropic  contribution per node is $\nu\ln
2$. This value is however only an  upper bound (known as the
annealed, or first moment bound) on the  entropy density $s=S_N/N$
that we wish to calculate. In fact, it holds only  if the bits are
independent: indeed, two bits may both be free but, by fixing
one, the second may be constrained to a unique value, in which
 case the joint entropic contribution of the two nodes is $\ln
2$ and  not $2\ln 2$. The correct expression is given by the Bethe
formula, which can be heuristically derived as follows.  First, we sum
the entropic contributions $\Delta  S_{\circ+\square\in\circ}$ of
each node $\circ$, including the corrections due to its adjacent
parity-checks $\square\in\circ$. Second, we note that each
parity-check $\square$ is involved in $k_\square$ terms, with
$k_\square$ being the connectivity of $\square$. To count it
only once, we therefore subtract $(k_\square-1)$ times the
entropic contribution $\Delta S_\square$ of each parity check
$\square$. This leads to
\begin{equation}
s=\frac{1}{N}\left(\sum_\circ\Delta
S_{\circ+\square\in\circ}-\sum_\square(k_\square-1)\Delta
S_\square\right)=\langle \Delta
S_{\circ+\square\in\circ}\rangle-\frac{\langle
\ell\rangle}{\langle  k\rangle}\sum_k c_k (k-1)\langle\Delta
S^{(k)}_\square\rangle
\end{equation}
where $\langle\Delta S_{\circ+\square\in\circ}\rangle$ represents
the  average of $\Delta S_{\circ+\square\in\circ}$ over the nodes
$\circ$,  and $\langle\Delta S^{(k)}_\square\rangle$ the average
of $\Delta  S_\square$ over the parity checks $\square$ with
connectivity  $k_\square=k$; the factor
$\langle\ell\rangle/\langle k\rangle$  accounts for the ratio of
the number $M$ of parity checks over the number $N$ of nodes.

\begin{figure}
\begin{center}
\epsfig{file=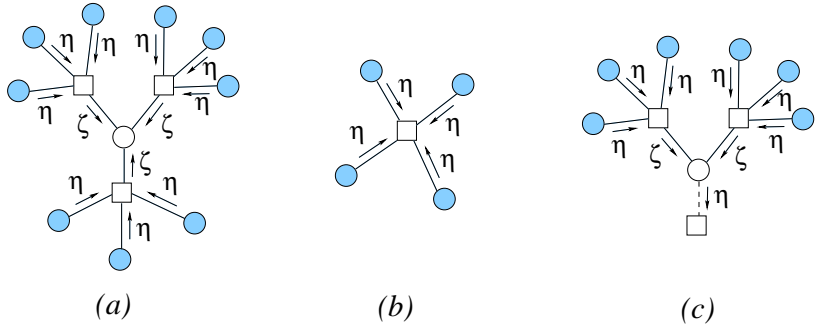,width=.6\linewidth}
\caption{\label{fig:cavity}\small Illustration of cavity fields:
(a) Addition of a variable node; (b) Addition of a parity check;
(c) Cavity iteration.}
\end{center}
\end{figure}

To compute $\Delta S_{\circ+\square\in\circ}$, we need to know
 whether the bits of the nodes adjacent to $\circ$ are fixed
or not, in  the absence of the ``cavity node'' $\circ$. As the
cavity node is connected to  its neighbors through parity checks
(see Fig.~\ref{fig:cavity}\,(a)), we can decompose the computation
in two steps. First, we observe that a given neighboring parity
check constrains the value of the cavity node if and only if all
the other nodes to which it is connected have themselves their bit
fixed {\it in the absence of the cavity node}. Denoting by $\zeta$
the probability of this event, and by $\eta$ the probability for a
node to be free {\it in the absence of one of its adjacent parity
check}, we thus have
\begin{equation}\label{cavzeta}
\zeta=\sum_k\frac{kc_{k}}{\langle
k\rangle}[1-(1-\eta)^{k-1}]=1-\frac{c'(1-\eta)}{\langle k\rangle},
\end{equation}
where $kc_{k}/\langle k\rangle$ is the probability for a parity
check be connected to $k-1$ nodes {\it in addition to the cavity
node} (see Fig.~\ref{fig:cavity}\,(a)) and $1-(1-\eta)^{k-1}$ is
the probability that at least one of these $k-1$ nodes is free in
the absence of the parity check.  Next, we observe that the
probability for the cavity node to be free is the probability that
none of its adjacent parity checks is constraining, that is
\begin{equation}\label{nuzeta}
\nu=p\sum_{\ell}v_\ell\zeta^\ell=p v(\zeta).
\end{equation}

In order to close the equations, we also need the probability for
the cavity node to be free in the absence of one of its connected
parity check (see Fig.~\ref{fig:cavity}\,(c)), which is
\begin{equation}\label{caveta}
\eta=p\sum_\ell\frac{\ell v_{\ell}}{\langle
\ell\rangle}\zeta^{\ell-1}=p\frac{v'(\zeta)}{\langle\ell\rangle},
\end{equation}
where $\ell v_{\ell}/\langle\ell\rangle$ represents the probability
for a node to be connected to $\ell-1$ parity checks in addition
to the one ignored.  The ``cavity fields'' $\eta$ and $\zeta$,
determined by~\eqref{cavzeta} and \eqref{caveta}, contain all the
information needed to evaluate the entropy. Thus $\langle\Delta
S_{\circ+\square\in\circ}\rangle$ is given by
\begin{equation}
\langle\Delta S_{\circ+\square\in\circ}\rangle=(\ln2)\left[p\
v(\zeta)-\langle \ell\rangle \zeta \right].
\end{equation}
The first term, $(\ln 2)pv(\xi)$ corresponds to $(\ln 2)\nu$, see
Eq.~\eqref{nuzeta}, the average entropic contribution of a node
$\circ$, and the second, $-(\ln 2)\langle \ell \rangle \zeta$,
subtracts the entropic reductions of its adjacent parity-check
nodes; indeed they are $\langle\ell\rangle$ in average, and each
is constraining the cavity node with probability $\zeta$.
Similarly, the average entropic reduction due to a parity-check
alone is
\begin{equation}
\langle\Delta S^{(k)}_\square\rangle=-(\ln 2)\left[1-(1-\eta)^k\right]
\end{equation}
since $1-(1-\eta)^k$ is the probability that at least one of the
$k$ connected nodes is free in the absence of the parity check
(see Fig.~\ref{fig:cavity}\,(b)). To sum up, the entropy is
determined by the formul\ae
\begin{equation}
s=(\ln 2)\left[pv\left(1-\frac{c'(1-\eta)}{\langle
k\rangle}\right)-\frac{\langle \ell\rangle}{\langle
k\rangle}\left(1-c(1-\eta)-\eta c'(1-\eta)\right)\right],
\end{equation}

\begin{equation}\label{cavity}
\eta=p\frac{v'\left(1-c'(1-\eta)/\langle k
\rangle\right)}{\langle\ell\rangle}.
\end{equation}

\begin{figure}
\begin{center}
\resizebox{.5\linewidth}{!}{\input{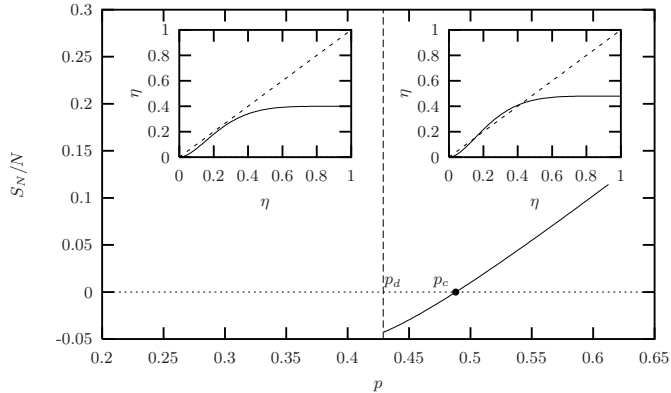}}
\end{center}
\caption{\label{fig:typical}Reduced entropy {\it vs.} noise level $p$ for an LDPC code with $k=6$ and $\ell=3$. When $p=0.4<p_d$ (left inset), $\eta=0$ is the only solution to the cavity equation \eqref{cavity}, yielding $s=0$. When $p=0.48>p_d$ (right inset), two more solutions appear, one of which is stable. The entropy of this solution crosses zero at the critical noise $p_c$, above which the entropy become strictly positive, causing failure of decoding.
}
\end{figure}

Eq.~\eqref{cavity} can admit two kinds of solution (see Fig.~\ref{fig:typical}). The
first kind, referred to as {\it ferromagnetic}, describes the
situation where decoding is possible, with only one codeword being
solution of the decoding-CSP: this solution has $\eta=0$ (all bits
are fixed to $\s^*$) and $s=0$. The second kind, referred to as
{\it paramagnetic} (but strictly speaking corresponding to a 1RSB glassy solution)
describes the situation where decoding is
impossible, and has $\eta>0$. It is found to exist only for $p$
greater than the so-called {\it dynamical threshold}, denoted by
$p_d$. It is however relevant only when associated with a positive
entropy, $s>0$, a condition which defines the {\it static
threshold}, denoted by $p_c$ and satisfying $p_c>p_d$. The
static threshold corresponds to the threshold above which decoding
is doomed to fail, as confirmed by rigorous studies. 

\subsubsection{Algorithmic interpretation}\label{sec:BP}

The cavity method is related to
a particular decoding algorithm known as belief propagation (BP).
Its principle is the following: starting from a configuration
where only the non-corrupted bits are fixed to their values, one
goes through each node of the factor graph, checks if its
immediate neighboring environment constrains it to a unique
value, fixes it to this value if it is the case, and iterates the whole
procedure until convergence. At the end, some bits may still not
be fixed, which certainly occurs if the decoding-CSP has not a
unique solution, but if all the bits end up fixed, one is insured
to have correctly decoded. Similar message-passing algorithms can
be defined with different channels. They are responsible for the
practical interest of LDPC codes as they provide algorithmically
efficient decoding (yet suboptimal, as discussed below). With the
BEC, these algorithms are particularly easy to analyze thanks to
the fact that one can never be fooled by fixing bits to an
incorrect value.     To perform the analysis of the possible
outcomes of the belief propagation algorithm, we can assume
without loss of generality that the transmitted message is
$(0,\ldots,0)$ (the $\Z_2$-symmetry implies that all codewords are
equivalent). We thus start with $\s_i=*$ if $i\in\mathcal{E}$, and
$\s_i=0$ otherwise. Cavity fields are attributed to each oriented
link of the factor graphs and are updated with the following
rules, where $t$ indexes iteration steps,
\begin{equation}\label{bpbec}
\begin{split}
h^{(t+1)}_{i\vers a}&=\left\{\begin{array}{rl}0 &\textrm{if $\s_i=0$ or if }u^{(t)}_{b\vers i}=1\textrm{ for some } b\in i-a \\ * &\textrm{otherwise}\end{array}\right.\\
u^{(t+1)}_{a\vers i}&=\left\{\begin{array}{rl}1 &\textrm{if }
h^{(t)}_{j\vers a}=0\textrm{ for all }j\in a-i \\ *
&\textrm{otherwise}\end{array}\right.
\end{split}
\end{equation}

Here, $u^{(t)}_{a\vers i}=1$ (resp. $*$) means that the parity
check $a$ is constraining (resp. is not constraining) $i$.
$h^{(t)}_{i\vers a}=0$ (resp. $*$) means that $\s_i$ is fixed
(resp. not determined) to its correct value 0 without taking into
account the constraints due to $a$. The algorithm is analyzed
statistically by introducing
\begin{equation}
\eta^{(t)}=\frac{1}{\langle \ell\rangle N}\sum_{(i,a)}
\delta(h^{(t)}_{i\vers a},0),\qquad \zeta^{(t)}=\frac{1}{\langle
k\rangle M}\sum_{(i,a)} \delta(u^{(t)}_{a\vers i},1).
\end{equation}

As suggested by our notations, the evolution for these quantities
exactly mimics the derivation of the formul{\ae} for the cavity
fields, yielding
\begin{equation}
\eta^{(t+1)}=p\frac{v'(\zeta^{(t)})}{\langle\ell\rangle},\qquad
\zeta^{(t+1)}=1-\frac{c'(1-\eta^{(t)})}{\langle k\rangle}.
\end{equation}

The fixed point is given by Eq.~\eqref{cavity}. When $p<p_d$,
the algorithm converges towards the unique, ferromagnetic, fixed
point $\eta^{(\infty)}=\zeta^{(\infty)}=0$, and decoding is
successfully achieved. When $p_d<p<p_c$, a paramagnetic
fixed point appears in addition to the ferromagnetic fixed point
and the iteration leads to this second
paramagnetic fixed point. The belief propagation algorithm thus
fails to decode above the dynamical threshold $p_d$, before reaching the
static threshold $p_c$ below which no algorithm can possibly be
successful (in this sense, BP is suboptimal).


\subsection{Average error exponents}\label{avrgbec}

\subsubsection{Entropic (1RSB) large deviations}

The previous section recalled the properties of typical codes subject to typical
noise. With finite codewords, $N<\infty$, failure to decode may also be due to atypical 
noise with unusually destructive effects. This is the purpose of our
large deviation approach to investigate such events. We first focus on
the simplest case, namely the computation of the average
error exponent where both the codes $\C$ and the noise $\xi$ are
treated on the same footing (see Sec.~\ref{sec:overview}). Our
procedure to deal with the statistics over atypical factor graphs is
an application of the cavity method for large deviations proposed in~\cite{Rivoire05}.
For the sake of simplicity, we restrain here to regular codes,
where nodes and check nodes have both fixed connectivity, $\ell$
and $k$ respectively, and defer the generalization to irregular
codes to Appendix~\ref{sec:irrdetails}.   

As explained in~\ref{sec:overview}, the thermodynamic formalism assigns a
Boltzmann weight $\e^{xS_N(\C,\xi)}$ to each ``configuration''
$(\C,\xi)$. The parameter $x$ plays the role of an inverse
temperature or, in other words, is a Lagrange multiplier enforcing
the value of $S_N$. Taking the infinite temperature limit $x=0$
(no constraint on the value of $S_N$) will thus lead us back to
the typical case discussed above.    

The cavity equations are as before derived by considering the effect of the addition of a
node. As adding a new node, along with its adjacent parity checks,
inevitably increases the degrees of the other nodes, strictly
restraining to regular graphs is not possible and we must work in
a larger framework. Accordingly, we consider ensembles where the
degree of parity checks is fixed to $k$, but where the degree of
nodes has a distribution $\{v_L\}$ (meaning that
degree $L$ has probability $v_{L}$, independently for each node).
We will describe the regular ensemble by taking
$v_L=\delta_{\ell,L}$ in the final formul\ae. Adding a new node with
$\ell$ parity-checks brings us from an ensemble characterized by
$v_L$ to an ensemble characterized by $v'_L$, with
\begin{equation}\label{degreeshift}
v'_{L}=\left(1-\frac{\ell(k-1)}{N}\right)v_L+\frac{\ell(k-1)}{N}v_{L-1}=v_L+\frac{\ell(k-1)}{N}\delta v_L
\end{equation}
where $\delta v_L=v_{L-1}-v_L$, since $\ell(k-1)$ nodes have their
degree increased by one. Let denote by $L(s,\{v_L\})$ the rate
function for the probability to observe $S_N/N=s$ in an ensemble
characterized by $\{v_L\}$, that is
\begin{equation}
\P_N[(C,\xi):S_N(C,\xi)/N=s\ |\ \{v_{L}\}]\asymp e^{-NL(s,\{v_{L}\})}.
\end{equation}
We introduce $P_{\circ+\square\in\circ}^{(\ell)}(\Delta S)$, the
probability distribution of the entropy contribution caused by the
addition of the new nodes along with its $\ell$ adjacent
parity-checks. The passage from $N$ nodes to $N+1$ nodes can then
be described by
\begin{equation}
\begin{split}
\P_{N+1}(s=S/(N+1)|\{v_{L}\}) &\asymp\ e^{-(N+1)L(S/(N+1),\{v_{L}\})}\\
&=\sum_{\ell}v_{\ell}\int \ud \Delta S\, P_{\circ+\square\in\circ}^{(\ell)}(\Delta S)\,\P_N[s=(S-\Delta S)/N|\{v_{L}-\ell(k-1)/N \delta v_L\}]\\
&\asymp\sum_{\ell}v_{\ell}\int \ud \Delta S\, P_{\circ+\square\in\circ}^{(\ell)}(\Delta S)\,\e^{-NL[(S-\Delta S)/N,\{v_{L}-\ell(k-1)/N \delta v_L\}]}.
\end{split}
\end{equation}
Expanding for large $N$, one gets
\begin{equation}\label{phi}
\phi_s(x)=xs-L(s,\{v_L\})=\ln\sum_{\ell}v_{\ell}\int \ud \Delta S\,
P_{\circ+\square\in\circ}^{(\ell)}(\Delta S)\,e^{x\Delta
S+z\ell(k-1)}
\end{equation}
with
\begin{equation}
z=\sum_{L}\delta v_L \frac{\partial L(s,\{v_{L}\})}{\partial v_{L}}.
\end{equation}
The parameter $z$ is determined by noting that the addition of a
new parity-check changes the node degree distribution in the same
way as in Eq.~\eqref{degreeshift}, with $v'_L=v_L+(k/N)\delta
v_L$, yielding
\begin{equation}
\e^{-NL(S/N,\{v_{L}\})}\asymp\int \ud\Delta S\,P_\square(\Delta
S)\, e^{-NL[(S-\Delta S)/N,\{v_L-(k/N)\delta v_L\}]},
\end{equation}
where $P_\square(\Delta S)$ is the probability of the entropy
reduction caused by the addition of a new parity-check. Expanding
here also for large $N$ leads to an equation for $z$,
\begin{equation}
z=-\frac{1}{k}\ln\int \ud\Delta S\,P_\square(\Delta S)\, e^{x\Delta S}.
\end{equation}

Following the same line of reasoning as in the typical case, the
two distributions $P_{\circ+\square\in\circ}^{(\ell)}$ and
$P_\square$ can be expressed by means of cavity fields $\eta$ and
$\zeta$. First consider the addition of a node: If the bit of the
new node is fixed, either because it was not erased or because one
its adjacent parity-check constrains it, there is an entropic
reduction $-\ln 2$ per non constraining adjacent parity-check, and
thus a weight $2^{-x}$. Otherwise, if the new node is free, which
occurs with probability $p\zeta^\ell$, the entropy shift is $(\ln
2)(1-\ell)$, giving a weight $2^{x(1-\ell)}$. Taking
$v_L=\delta_{L,\ell}$, Eq.~\eqref{phi} therefore reads
\begin{equation}\label{calcphi}
\phi_s(x)=\ln\left[{\left(\zeta 2^{-x}+1-\zeta\right)}^\ell-p(\zeta
2^{-x})^\ell+p\zeta^\ell 2^{x(1-\ell)}\right]+\ell(k-1)z,
\end{equation}
with
\begin{equation}\label{defzeta}
\zeta=1-(1-\eta)^{k-1}.
\end{equation}
Similarly, a new parity-check removes a degree of freedom if and
only if one of its adjacent node is free, which happens with
probability $1-(1-\eta)^k$, yielding
\begin{equation}\label{calcz}
z=-\frac{1}{k}\ln\left[1-(1-(1-\eta)^k)+(1-(1-\eta)^k)2^{-x}\right].
\end{equation}
Finally, we obtain a self-consistent equation for $\eta$ by
considering the addition of a new (cavity) node in the absence of
one of its adjacent parity-checks:
\begin{eqnarray}
\eta&=&\P(\textrm{cavity node free})\propto\int \ud \Delta S \, P_{\circ\vers\square}(\Delta S|\textrm{cavity node free}) e^{x\Delta S+z(\ell-1)(k-1)},\\
1-\eta&=&\P(\textrm{cavity node fixed})\propto\int \ud \Delta S \,
P_{\circ\vers\square}(\Delta S|\textrm{cavity node fixed})
e^{x\Delta S+z(\ell-1)(k-1)}
\end{eqnarray}
where $P_{\circ\vers\square}$ corresponds to
$P_{\circ+\square\in\circ}^{(\ell-1)}$, taken either under the
condition that the cavity node is free, or that it is fixed. We
obtain:
\begin{equation}\label{calceta}
\eta=\frac{p2^x\left(\zeta 2^{-x}\right)^{\ell-1}}{{\left(\zeta
2^{-x}+1-\zeta\right)}^{\ell-1}+p(2^{x}-1)\left(\zeta
2^{-x}\right)^{\ell-1}}.
\end{equation}
Alternatively, these equations can be obtained by differentiation of Eq.~\eqref{calcphi}, which is variational with respect to the cavity $\eta$.
The large deviation cavity equations~\eqref{defzeta} and
\eqref{calceta} allow us to compute the generating function
$\phi_s(x)$ using Eq.~\eqref{calcphi} and \eqref{calcz}, from which
the rate function $L(s|\{v_l=\delta_{l,\ell}\})$ is deduced by
Legendre transformation as discussed in~\ref{sec:overview}. 

\begin{figure}
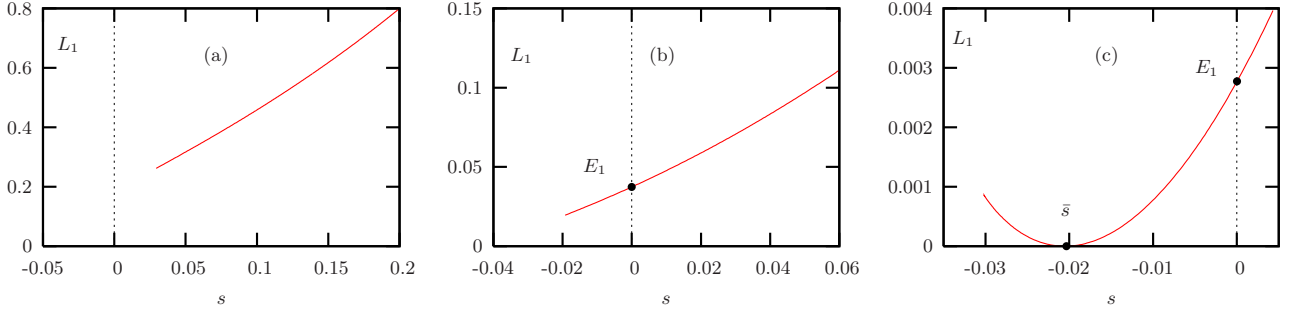

\begin{center}
\small
\resizebox{.32\linewidth}{!}{\input{largedevbec1}}
\resizebox{.32\linewidth}{!}{\input{largedevbec2}}
\resizebox{.32\linewidth}{!}{\input{largedevbec3}}
\caption{\label{beclargedev}\small Rate function ${L}(s)$ as
a function of the entropy $s$, here illustrated with a regular
code with $k=6$ and $\ell=3$ (for the BEC channel). The three
regimes are represented. (a) $p=0.2<p_{\textrm{1RSB}}$: the spinodal of the
paramagnetic solution is for $s_d>0$. (b) $p=0.35\in[p_{\textrm{1RSB}},p_d]$:
the spinodal is now for $s_d<0$. (c) $p=0.45\in [p_d,p_c]$: the
spinodal is preceded by a minimum (the typical value), with
$x_d=\partial_sL(s=s_d)<0$. The typical dynamical and static
transitions can be read on the $s=0$ axis: by definition of
$p_d$ and $p_c$, this equation has a solution $\bar s$ for
$p>p_d$, and this solution is positive, $\bar s>0$, for
$p>p_c$ (not represented here).}
\end{center}
\end{figure}

Again, two kinds of solutions, paramagnetic or ferromagnetic, can be present.
For a given value of $p$, we find that a non-trivial, paramagnetic
solution to Eq.~\eqref{calceta} exists only for $x\geq x_d(p)$. In
agreement with the observation reported in the previous section
that the paramagnetic solution typically exists only when
$p<p_d$, we have $x_d(p)<0$ for $p>p_d$ and $x_d(p)>0$
for $p<p_d$ (the typical case is indeed associated with $x=0$). We obtain the average error exponent by selecting the value of $L(s)$ where $s=0$: our results are illustrated in Fig~\ref{beclargedev}.
By extension of the concept of dynamical threshold $p_d$, one could define a ``dynamical'' error exponent as $E_d(p)={L}(x_d(p))=x_d(p)s(x_d(p))-\phi(x_d(p))$ with $x_d(p)$ corresponding to the temperature of the spinodal for the paramagnetic solution. The relevance of this concept is however limited by the fact that the algorithmic interpretation presented in Sec.~\ref{sec:BP} does not extend to large deviations (see also Sec.~\ref{sec:notealgo}).

 More interestingly, we find an additional threshold, denoted $p_{1RSB}$, below which the equation $s(x)=0$ has no longer a solution (see Fig.~\ref{beclargedev}). This inconsistency of the 1RSB solution is indicative of the presence of a phase transition occurring at some $p_e>p_{\rm 1RSB}$. The following section is devoted to computing $p_e$, and describing the nature of the new phase present for $p<p_e$.

\begin{figure}
\begin{center}
\resizebox{.5\linewidth}{!}{\input{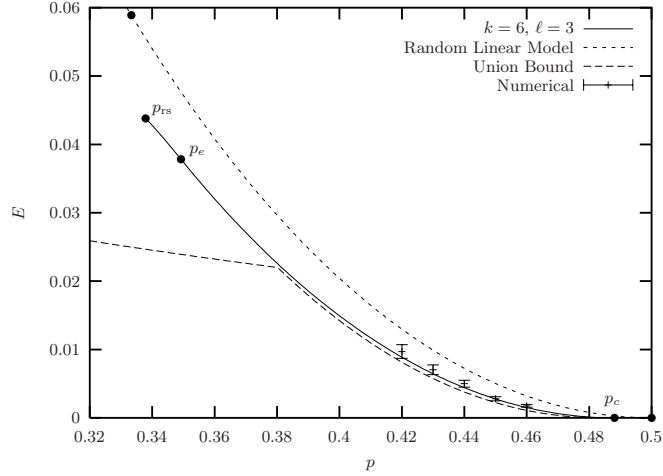}}
\caption{\label{becerrexp}\small Average error exponent as
a function of the noise level $p$ for the regular code ensemble
with $k=6$ and $\ell=3$, on the BEC. Numerical estimates of the error probability, based on $10^6$ runs of exact Maximum Likelihood decoding (using Gauss elimination) on samples of sizes ranging from $N=500$ to $N=1500$, yield reasonably good estimates of the error exponent using an exponential fit. These numerical results agree well with our theoretical prediction.
The union bound \eqref{unionbec} and
the random linear limit \eqref{RLM} are also represented for comparison.}
\end{center}
\end{figure}
\begin{table}
\begin{center}
\begin{tabular}{|l|l|l|l|l|l|}
\hline  $(k,\ell)$ & $p_{\textrm{1RSB}}$ & $p_{\rm RS}$ & $p_e$ & $p_d$ & $p_c$ \\  \hline  $(4,3)$ &
$0.3252629709$ & $0.5465748811$ & $0.6068720166$ &$0.6474256494$ & $0.7460097025$
\\  \hline  $(6,3)$ & $0.2668568754$ & $0.3378374641$ & $0.3491884902$ &
$0.4294398144$ & $0.4881508842$ \\  \hline  $(6,5)$ &
$0.01300820524$ & $0.4277010368$ & $0.7143657513$ & $0.5510035344$ & $0.8333153204$ \\ \hline
$(10,5)$ & $0.04412884546$ & $0.2435656894$ & $0.3347721176$ & $0.3415500230$ & $0.4994907179$
\\  \hline

\end{tabular}
\caption{\label{noiselevels}\small Values of some thresholds
$p_{\textrm{1RSB}}$, $p_{\rm RS}$, $p_e$, $p_d$ and $p_c$ for different regular ensembles of LDPC codes on the BEC.}
\end{center}
\end{table}

\subsubsection{Energetic (RS) large deviations}

The previous ``entropic (1RSB) approach'' attributed errors to the presence of an exponential number of solutions in the decoding-CSP. The same assumption was underlying the analysis of the typical case, in Sec.~\ref{sec:typBEC}, where rigorous studies support the conclusions drawn from this hypothesis. This view is also consistent with the phase diagram of XORSAT problems to which the encoding-CSP belongs. The structure of the well separated codewords corresponds in this context to a ``frozen 1RSB glassy'' phase. As $p$ departs from the value $p=1$ however, the decoding-CSP deviates increasingly in nature from the initial encoding-CSP. As the number of constraints increases (as $p$ decreases), the presence of an exponential number of solutions (glassy phase) in addition to the isolated correct codeword becomes less and less probable. An alternative rare event possibly dominating the probability of error at low $p$ is the presence of a second isolated (ferromagnetic) codeword close to the correct one. This can lead to a new phase transition that has no counterpart in the typical phase diagram, reflected by a non-analyticity of the error exponent.

In our framework, investigating an alternative source of error requires considering for $S_N$ an other quantity than the entropy of the number of solutions. A possible choice, associated with a replica symmetric (RS) Ansatz, is the energy $E_N$ of the ground-state of the decoding-CSP, giving the minimal number of violated parity checks. Ignoring the correct codeword, a second isolated codeword is present if and only if $E_N=0$ (otherwise $E_N>0$). Large deviations of this energy are described by the rate function $L_1(\e)$ defined as
\begin{equation}
\P[\xi,\C:E_N(\xi,\C)/N=e]\asymp \e^{-NL_1(e)}.
\end{equation}
The generating function for the rate function $L_1(\e)$, defined by
\begin{equation}
e^{N\phi_e(x)}=\E_{\xi,\C}\left[\e^{xE_N(\xi,\C)}\right]=\int \ud e\,
\e^{N(xe-L_1(e))}.
\end{equation}
is given by (see~\cite{MezardRicci03} for a similar calculation)
\begin{equation}
\begin{split}
\phi_1(x)&=\ln\left[p\int\prod_{a=1}^{\ell}\ud u_a\,Q(u_a)\e^{-x\left(\sum_{a=1}^{\ell}|u_a|-|\sum_{a=1}^{\ell}u_a|\right)}+(1-p)\int\prod_{a=1}^{\ell}\ud u_a\,Q(u_a)\e^{-2x\sum_{a=1}^{\ell}\delta_{u_a,-1}}\right]\\
&-\frac{\ell(k-1)}{k}\ln\int\prod_{i=1}^{k}\ud h_i\,
P(h_i)e^{-x\delta(\prod_{i=1}^{k}h_i,-1)}
\end{split}
\end{equation}
with
\begin{eqnarray}
P(h\neq+\infty)&\propto &p\int\prod_{a=1}^{\ell-1}\ud u_a\,Q(u_a)\e^{-\frac{x}{2}\left(\sum_{a=1}^{\ell-1}|u_a|-|\sum_{a=1}^{\ell-1}u_a|\right)}\delta\left(h-\sum_{a=1}^{\ell-1}u_a\right)\\
P(h=+\infty)&\propto & (1-p)\int\prod_{a=1}^{\ell-1}\ud u_a\,Q(u_a)\e^{-x\sum_{a=1}^{\ell-1}\delta_{u_a,-1}}\\
Q(u)&=&\int\prod_{i=1}^{k-1}\ud h_i\,
P(h_i)\delta\left[u-\mathcal{S}\left(\prod_{i=1}^{k-1}h_i\right)\right]
\end{eqnarray}
where $\mathcal{S}(x)=1$ if $x>0$, $-1$ if $x<0$, and $0$ if
$x=0$. Since $u$ only takes values in $\{-1,0,+1\}$, and $h$ is restrained to integer values, we can introduce
\begin{equation}
Q(u)=q_+\delta(u-1)+q_-\delta(u+1)+q_0\delta(u),
\end{equation}
and
\begin{equation}
p_+=\int_{h>0} \ud h\,P(h)\qquad p_-=\int_{h<0} \ud
h\,P(h)\qquad p_0=1-p_+ - p_-.
\end{equation}
Our interest is here in zero-energy ground states, described by the limit $x\vers \infty$ where the equations simplify to:
\begin{equation}
\phi_\e(x=+\infty)=-L(e=0)=\ln\left[(1-q_-)^{\ell}+p(1-q_+)^{\ell}-p
q_0^{\ell}\right]-\frac{\ell(k-1)}{k}\ln\left[1-\frac{1}{2}\left((p_++p_-)^k-(p_+-p_-)^k\right)\right],
\end{equation}
with
\begin{eqnarray}
p_+&\propto& (1-q_-)^{\ell-1}-p q_0^{\ell-1},\\
p_-&\propto& p (1-q_+)^{\ell-1}-p q_0^{\ell-1},\\
p_0&\propto& p q_0^{\ell-1},\\
q_+&=& \frac{1}{2}\left[(p_+ +p_-)^{k-1}+(p_+ - p_-)^{k-1}\right],\\
q_-&=& \frac{1}{2}\left[(p_+ +p_-)^{k-1}-(p_+ - p_-)^{k-1}\right],\\
q_0&=& 1-(p_+ +p_-)^{k-1}.
\end{eqnarray}
We find that the only stable solution to these cavity equations
satisfies $q_0=p_0=0$, which allows us to further simplify the formul{\ae},
\begin{equation}\label{eq:phiinf}
\phi_\e(+\infty)=\ln\left[q_+^\ell+p(1-q_+)^\ell\right]-\frac{\ell(k-1)}{k}\ln\left[\frac{1}{2}(1+(2p_+-1)^k)\right],
\end{equation}
with
\begin{eqnarray}\label{eq:cavRS1}
p_+&=&\frac{q_+^{\ell-1}}{q_+^{\ell-1}+p(1-q_+)^{\ell-1}},\\
q_+&=&\frac{1}{2}\left[1+(2p_+-1)^{k-1}\right].\label{eq:cavRS2}
\end{eqnarray}
The resulting RS average error exponent, given by $E_e(p)=-\phi(+\infty)$, is
represented in Fig.~\ref{becerrexp}. 

We identify the transition $p_e$ as the point where the 1RSB and RS error exponents coincide, which satisfies $p_e>p_{\rm 1RSB}$. We find that the RS solution is limited by a spinodal point and is only defined for $p\geq p_{\rm RS}$. While we conjecture that the 1RSB estimate is exact for $p>p_e$, the existence of $p_{RS}$ suggests that either an additional phase transition occurs at some $p_e'>p_{RS}$, or, more radically, that our description of the phase $p<p_e$ is incorrect. The limit case of random codes however indicates that the energetic method is valid in the limit $k$, $\ell\to\infty$.

\subsubsection{Limit of random codes}

 The only limiting case where the average error exponent has been obtained integrally so far is the fully connected limit where $k,\ell\vers\infty$ with $\ell/k=\alpha=1-R$ fixed. This limit corresponds to the {\it random linear model} (RLM), where each parity-check is connected to each node with probability $1/2$. In this limit, the entropic 1RSB approach gives
\begin{equation}\label{klinfty}
E_s(k,\ell\vers \infty)=L(s=0)=D(1-R||p),
\end{equation}
where $D(q||p)=q\ln(q/p)+(1-q)\ln((1-q)/(1-p))$ is known as the
Kullback-Leibler divergence, while the energetic RS approach gives
\begin{equation}
E_e(k,\ell\vers \infty)=-\phi_e(+\infty)=-(R-1)\ln 2-\ln(1+p),
\end{equation}
(with $p_+=1/1+p$ and $q_+=1/2$). The two expression coincide at the critical noise $p_e$, with
\begin{equation}
p_e=(1-R)/(1+R).
\end{equation}
We thus predict the average error exponent of the RLM to be
\begin{equation}\label{RLM}
E_{1}(\textrm{RLM})=\left\{\begin{array}{ll}(1-R)\ln 2-\ln(1+p) 
& \textrm{if }p<\frac{1-R}{1+R},\\
D(1-R||p)&\textrm{if }\frac{1-R}{1+R}<p<1-R.\end{array}\right.
\end{equation}
This result coincides with the exact expression (see Appendix~\ref{sec:RLM} for a direct combinatorial derivation), thus validating our approach in this particular case.

As explained above, we are not able to fully account for the small noise regime as soon as $k$ and $\ell$ are finite, even though the solutions are found to be stable with respect to further replica symmetry breakings in the space of codewords \cite{MontanariRicci03}. This does not exclude that a similar replica symmetry breaking occurs in the space of codes.
Remarkably, previous attempts reported in the literature have also failed to obtain error exponents in the low $p$ regime.


\subsection{Typical error exponents}

\subsubsection{Cavity equations}\label{sec:typcav}

The typical error exponent is encoded into a potential $\psi(x,y)$, as defined in Eq.~\eqref{eq:psidef}. The equations for $\psi(x,y)$ are obtained from the cavity method for large deviations by following very closely the path leading to $\phi(x)$~\cite{footnoteContrary}. As noticed in Sec.~\ref{sec:part2}, the formalism with finite $y$ provides a generalization of the average case which is recovered by taking $y=1$, with $\psi(x,y=1)=\phi(x)$. We will therefore only quote our results. In the entropic (1RSB) case, we find
\begin{equation}
\psi_s(x,y)=\ln\left[(\zeta 2^{-xy}+1-\zeta)^\ell-(\zeta 2^{-xy})^\ell+\zeta^\ell(p2^x+1-p)^y2^{-\ell xy}\right]-\frac{\ell(k-1)}{k}\ln\left[(1-\eta)^k+(1-(1-\eta)^k)2^{-xy}\right]
\end{equation}
with
\begin{equation}
\begin{split}
\eta &=\frac{\zeta^{\ell-1}(p2^x)^y2^{-(\ell-1)xy}}{(\zeta2^{-xy}+1-\zeta)^{\ell-1}-(\zeta2^{-xy})^{\ell-1}+\zeta^{\ell-1}(p2^x+1-p)^y2^{-(\ell-1)xy}},\\
\zeta &=1-(1-\eta)^{k-1}.
\end{split}
\end{equation}
In the energetic (RS) case with $x=+\infty$, we find
\begin{equation}
\psi_\e(x=+\infty,y)=\ln\left[q_+^\ell+p^y(1-q_+)^\ell\right]-\frac{\ell(k-1)}{k}\ln\left[\frac{1}{2}(1+(2p_+-1)^k)\right],
\end{equation}
with
\begin{eqnarray}
p_+&=&\frac{q_+^{\ell-1}}{q_+^{\ell-1}+p^y(1-q_+)^{\ell-1}},\\
q_+&=&\frac{1}{2}\left[1+(2p_+-1)^{k-1}\right].
\end{eqnarray}
In each case, from the potential $\psi(x,y)$, the rate function is obtained as $\mathcal{L}(\phi,x)=y\phi-\psi(x,y)$, with $\phi(x)=\partial_y\psi(x,y)$. By definition, a typical code corresponds to a minimum of $\mathcal{L}$, with $\mathcal{L}=0$, which, when $\mathcal{L}$ is analytical at this minimum, is associated with $y=\partial_\phi\mathcal{L}=0$. 

\begin{figure}
\begin{center}
\resizebox{.7\linewidth}{!}{\input{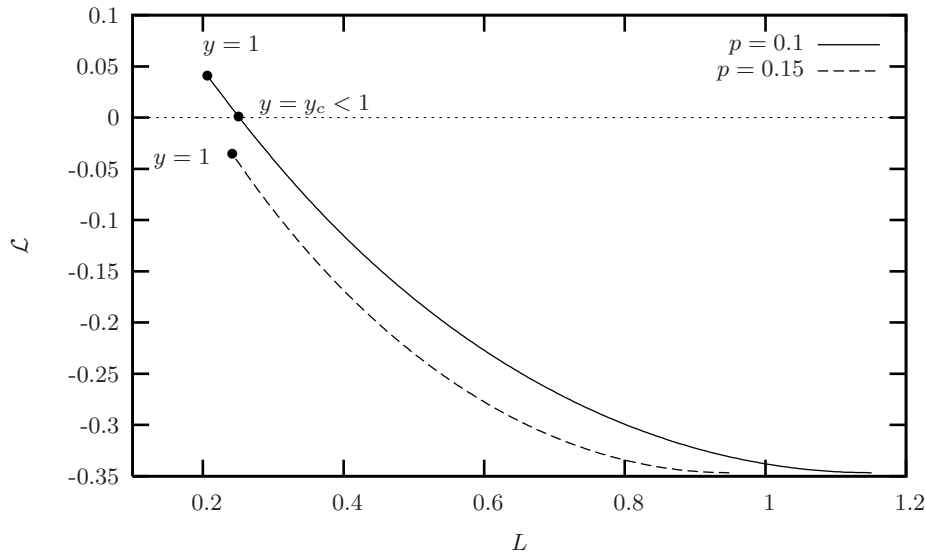}}
\caption{\label{fig:2stepbec}\small Rate function $\L(L_e)=\L[-\phi_e(+\infty)]$ of the energetic error exponent for an
 LDPC code with $k=24$, $\ell=12$ on the BEC. When $p>p_y$ (full curve), the rate function is negative (and therefore unphysical) for all $0<y<1$, entailing that the typical and average error exponent should coincide.
When $p<p_y$ (dashed curve), we postulate that the typical error exponent is given by the inverse ``freezing temperature'' $y_c$ at which the rate function cancels.}
\end{center}
\end{figure}

As a generic feature,  we find that $\mathcal{L}(y,x)$ is an increasing function of $y$ for fixed $x$, going from negative values for $y<y_c(x)$ to positive ones for $y>y_c(x)$. Negative rate functions, as thus obtained, are certainly unphysical. As negative entropies in the usual cavity/replica method, we attribute them to analytical continuations of physical solutions. 
The simplest way to circumvent them is, as with the frozen 1RSB Ansatz in the replica method, to select $y_c(x)$ with $\mathcal{L}(y,x)=0$. 
When $y_c(x)<1$, meaning that $\mathcal{L}(y=1,x)>0$, we consider that the average exponent is associated with atypical codes and therefore differs from the typical exponent, described by $\mathcal{L}(y_c(x),x)=0$. Using this criterion, we find that the two exponents indeed differ for the lowest values of $p$, when $p<p_y$, where $p_y<p_e$ (see Fig.~\ref{fig:2stepbec} for an illustration).
In general the situation is complicated by the fact that the cavity equations may fail to provide solutions in this regime, as already seen in the average case when $p<p_{\rm RS}$ (corresponding here to $y=1$); the random code limit, where this complication is absent, is thus the most instructive.

\subsubsection{Limit of random codes}

In the limit $k$, $\ell\to\infty$, we obtain the following results. In the entropic regime, $p>p_e$, the average and typical exponents are found to coincide. This conclusion extends in the energetic regime only for a restricted interval $[p_y,p_e]$. When $p<p_y$, we have $y_c(x)<1$ and average and typical error exponents differ. The formula we obtain for the typical error exponent reads
\begin{equation}
E_{\rm typ}(\textrm{RLM})=\left\{
\begin{array}{ll}-\delta_{GV}(R)\log p & \textrm{if }p<p_y,\\
E_{\rm av}(\textrm{RLM})&\textrm{if }p_y<p<p_c,
\end{array}\right.
\end{equation}
with
\begin{equation}\label{eq:py}
p_y=\frac{\delta_{GV}(R)}{1-\delta_{GV}(R)}.
\end{equation}
$\delta_{GV}(R)$ denotes the smallest solution to $(R-1)\ln 2+H(\delta)=0$, whose interpretation is discussed in Appendix~\ref{sec:RLM}. This result, which does not seem to have been reported previously in the literature, coincides with the union bound presented in Appendix~\ref{sec:union}, which strongly suggests that it is indeed exact.

For LDPC with finite connectivity, a similar phase diagram is expected. In the entropic regime, we find indeed that average and typical exponents are identical. In the energetic regime, we face the problem that the cavity equations have no solution below some value of $p$, which precludes us from estimating $p_y$.

\subsubsection{Algorithmic implications}\label{sec:notealgo}

The cavity formalism has the attractive property of corresponding formally to message passing algorithms. Based on this analogy, new algorithmic procedures have been systematically proposed to analyze single finite graphs, each time the cavity approach was found to operate at the ensemble level. With a phase transition occurring at the ensemble level, we have however here a system where such a correspondence is no longer meaningful. Following the usual procedure, it is indeed straightforward to implement the cavity approach for average error exponent on a single graph, but in the regime $p<p_y$, this algorithm is doomed to fail: for any typical graph, in the limit of large size, the message passing algorithm will yield the average error exponent, which, as we have seen, is distinct for the correct, typical, error exponent.


\section{LDPC codes over the BSC}

\subsection{Definition}

We now turn to error exponents for LDPC codes on the binary symmetric channels (BSC). One motivation for repeating the analysis with this channel is that it is representative of a broader class of channels, where bits are not simply erased as with the BEC, but can be {\it corrupted}, in the sense that their content 0 or 1 is changed to other admissible values. This clearly complicates the decoding as corrupted bits can not be straightforwardly identified; in fact, with the BSC, no scheme can guarantee to identify the corrupted bits, and the receiver is never certain that his decoding is correct. We will however see that the overall phase diagram is very similar to that obtained with the BEC.

By definition, maximum-likelihood decoding consists in inferring the most probable realization of the noise {\it a posteriori}. The {\it a posteriori} probability can be expressed from the {\it a priori} probability thanks to Bayes' theorem. If $\bx$ denotes the transmitted message and $\by$ the received message, the {\it a priori} probability to receive $\by$ given $\bx$ is
\begin{equation}
Q(\by|\bx)=\prod_{i=1}^N (1-p)^{\delta_{x_i,y_i}}p^{1-\delta_{x_i,y_i}}.
\end{equation}
To make contact with physical models of disordered systems \cite{Sourlas89}, it is convenient to adopt a spin convention: $\sigma_i=(-1)^{x_i}$, $\tau_i=(-1)^{y_i}$, and to rewrite the previous relation as
\begin{equation}
Q(\bs|\btt)\propto 
e^{\sum_{i=1}^N h_i\t_i},\qquad h_i\equiv h_0\s_i,\qquad
h_0 \equiv \frac{1}{2}\ln\left(\frac{1-p}{p}\right).
\end{equation}
This formulation emphasizes the analogy with the random field Ising model~\cite{Nattermann98}, a prototypical disordered system. Using the group symmetry of the set of codewords, we can assume, without loss of generality, that the sent codeword is $\bs=(+1,\ldots,+1)$. With this simplification, the random field takes value $h_i=h_0$ with probability $1-p$ and $-h_0$ with probability $p$. Bayes' formula for the {\it a posteriori} probability that the message $\btt$ was sent reads
\begin{equation}\label{hamiltonian}
P(\btt|\bs)=\frac{P(\bs|\btt)P(\btt)}{\sum_{\t'}P(\bs|\btt')P(\btt')}=\frac{1}{Z(\beta)}e^{\beta\sum_{i=1}^Nh_i\t_i}\prod_{a=1}^M\delta(\t_a=1)
\end{equation}
where $\t_a$ is a shorthand for $\prod_{i\in a}\t_i$: in the present spin convention, the constraint induced by the parity-check $a$ indeed reads $\t_a=1$. To continue the analogy with statistical mechanics, we have also introduced a temperature $\beta$, called the decoding temperature, whose value is here fixed to $\beta=1$ (Nishimori temperature ---see \cite{NishimoriBook01}). Given the {\it a posteriori} probability, the selection of the most probable codeword ${\bf d}(\bs)$ can still be done according to different criteria, amongst which:
\begin{itemize}
\item {Word maximum a posteriori (word-MAP)}, where one maximizes the posterior probability in block by taking ${\bf d}_{\textrm{block}}(\s)=\textrm{argmax}_{\btt}P(\btt|\bs)$. This scheme minimizes the block-error probability, $P_{\rm block}=(1/M)\sum_{\btt} \P[{\bf d}(\bs)\neq\bs]$.
\item{Symbol maximum a posteriori (symbol-MAP)}, where one maximizes the posterior probability bit per bit by taking ${\bf d}_{\textrm{bit}}(\bs)_i=\textrm{argmax}_{\t_i}\sum_{\t_{j\neq i}}P(\btt|\bs)$. This scheme minimizes the bit-error probability $P_{\rm bit}=(1/M)\sum_{\btt} (1/N)\sum_i \P[{\bf d}(\btt)_i\neq\s_i]$.
\end{itemize}
In physical terms, the word-MAP procedure consists in finding the ground state of the system with partition function $Z(\beta)$ given by the normalization in Eq.~\eqref{hamiltonian}; this amounts to studying the zero-temperature limit, $\beta\vers \infty$. Conversely, symbol-MAP is equivalent to taking the sign of the local magnetizations at temperature $\beta=1$,
\begin{equation}
\t^{\textrm{bit}}_i=\textrm{sign}\left(\langle \t_i\rangle\right)=\textrm{sign}\left[\sum_{\btt}\t_i P(\btt|\bs)\right].
\end{equation}
We will treat the two cases in a common framework by considering an arbitrary temperature $\beta\geq 1$.

From the physical perspective, the original codeword is recovered if it dominates the Gibbs measure defined in Eq.~\eqref{hamiltonian}. This can be expressed by decomposing the partition function $Z(\beta)$ as
\begin{equation}
Z(\beta)=Z_{\textrm{corr}}(\beta)+Z_{\textrm{err}}(\beta),\qquad Z_{\textrm{corr}}(\beta)=\e^{\beta\sum_i h_i},\qquad Z_{\textrm{err}}(\beta)=\sum_{\btt\neq \mathrm{1}}\e^{\beta\sum_i h_i\t_i}\prod_a\delta(\tau_a-1).
\end{equation}
We define the corresponding free energies, $F_{\textrm{corr}}(\beta)=-(1/\beta)\ln Z_{\textrm{corr}}(\beta)$ and $F_{\textrm{err}}(\beta)=-(1/\beta)\ln Z_{\textrm{err}}(\beta)$. The first one corresponds physically to a ferromagnetic phase (as with the BEC), while the second will be shown to correspond either to a paramagnetic or a glassy phase, depending on the values of $\beta$ and $p$. Decoding is successful if, and only if, the ferromagnetic phase has lower free energy, $F_{\rm corr}<F_{\rm err}$. The quantity $S_N(\xi,\C)$ introduced in section~\ref{sec:overview} can therefore be defined here as
\begin{equation}
S_N=F_{\textrm{corr}}(\beta)-F_{\textrm{err}}(\beta)
\end{equation}
where the dependence in the noise $\xi$ and the code $\C$ is implicitly understood.

\subsection{Cavity analysis and the 1RSB frozen ansatz}

As with the BEC, explicit calculations can be performed by means of the replica or cavity methods. Details can be found in Appendix~\ref{sec:BSCdetails} and we only discuss here the points where differences with the BEC arise. For any fixed $p$, a replica symmetric (RS) calculation, whose derivation follows the derivation of the paramagnetic solution with the BEC, is found to undergo an entropy crisis, i.e., $s_{\rm RS}(\beta)=\beta^2\partial_\beta f_{\rm RS}(\beta)<0$ for $\beta>\beta_g$. This feature is indicative of the presence of a glassy phase, and points to the need to break the replica symmetry. The glassy phase of LDPC codes is however of the ``frozen 1RSB'' type, which implies that the glassy free energy $f_{\rm err}$ can be completely inferred from the replica symmetric solution $f_{\rm RS}$. This simplicity stems from the ``hard'' nature of the constraints: changing a bit automatically violates all its surrounding checks, forcing the rearrangement of many variables \cite{MontanariSemerjian05,MontanariSemerjian05-2}. When the degree of all nodes is $\ell_i\geq 2$, one can indeed show~\cite{MezardRicci03} that changing one bit while keeping all checks satisfied requires the rearrangement of an extensive ($\propto N$) number of variables (in the language of~\cite{MezardRicci03}, factor graphs of LDPC codes have no leaves). The consequence, expressed in the replica language, is that the 1RSB ``states'' are reduced to single configurations, and thus have zero internal entropy. The 1RSB potential $\phi(\beta,m)$ whose optimization over $m\in[0,1]$ is predicted to yield $f_{\rm err}$~\cite{MezardParisi87b} thus simplifies to $\phi(\beta,m)=f_\textrm{RS}(\beta m)$~\cite{MartinMezard05}, since
\begin{equation}
\e^{-N\beta m\phi(\beta,m)}\equiv \sum_{\textrm{states }\alpha}e^{-N\beta mf_\alpha(\beta)}=\sum_{\alpha}e^{-N \beta me_\alpha}=\e^{-\beta m f_{\rm RS}(\beta m)}.
\end{equation}
According to whether one is above or below the freezing temperature $\beta_g^{-1}$, defined by
\begin{equation}
s_{\rm RS}(\beta_g)=\beta_g^2\partial_\beta f_\textrm{RS}(\beta_g)=0,
\end{equation}
the free energy $f_{\rm err}(\beta)$ is given either by $f_\textrm{RS}(\beta)$ (paramagnetic phase), or by $f_\textrm{RS}(\beta_g)$ (glassy phase). This is summarized as follows:
\begin{equation}\label{fpara}
f_\textrm{err}(\beta)=\max_{\beta'<\beta}f_\textrm{RS}(\beta')=\left\{\begin{array}{ll} f_\textrm{RS}(\beta) &\textrm{if }\beta<\beta_g, \\
f_\textrm{RS}(\beta_g) & \textrm{if }\beta>\beta_g.\end{array}\right.
\end{equation}

Finally, we note that as in the BEC case, a non-ferromagnetic solution $f_\textrm{RS}(\beta)$ exists only for large enough $p$. The threshold $p_d(\beta)$ giving the smallest noise level at which a non-ferromagnetic solution exists is again called the dynamical threshold, and can be shown here also to coincide with the dynamical arrest of BP~\cite{FranzLeone02}.

\subsection{Average error exponent}\label{sec:avrgbsc}
\subsubsection{LDPC codes}

In the region relevant for error exponents, where $p<p_c$ and $\beta\geq 1$ , the ferromagnetic solution is typically dominant (this is the definition of $p<p_c$), and metastable phases described by $f_{\rm err}$ are typically glassy, since $\beta_g<1$. Therefore, to compute error exponents, we have to consider $f_{\rm err}(\beta)=f_{\rm RS}(\beta_g)$, and not $f_{\rm err}(\beta)=f_{\rm RS}(\beta)$. This leads us to introduce an extra temperature $\beta_e$ distinct from the decoding temperature $\beta$, which is to be set to $\beta_g$ by requiring that the entropy $s_{\rm RS}$ is zero. Similarly, we introduce a ferromagnetic temperature $\beta_f$, set to $\beta_f=\beta$, and define the rate function $L_1(f_e,f_f)$ and its Legendre transform as
\begin{equation}\label{eq:ldcmbsc}
\begin{split}
&\P[\xi,\C:F_\textrm{RS}(\beta_e)/N=f_e,F_\textrm{corr}(\beta_f)/N=f_f]\asymp \e^{-NL_1(f_e,f_f)},\\
&\e^{N\phi_1(\beta_e,\beta_f,x_e,x_f)}=\E_{\xi,\C}\left[\e^{-x_e \beta _e F_{\textrm{RS}}(\beta_e)-x_f \beta _f F_{\textrm{corr}}(\beta_f)}\right]=\int \ud f_e\,\ud f_f\, \e^{N\left[-x_e \beta_e f_e-x_f \beta_f f_f-L_1(f_e,f_f)\right]}
\end{split}
\end{equation}
The potential $\phi_1$ contains all the necessary information about both solutions:
\begin{equation}\label{freeen}
-\beta_af_a=\partial_{x_a}\phi_1,\qquad
s_a=\partial_{x_a}\phi_1-\frac{\beta_a}{x_a}\partial_{\beta_a}\phi_1,
\end{equation}
where the index $a=e,f$ corresponds to the two possible phases. 
To the purpose of computing error exponents, we need only to control $f_e-f_f$ and $s_e$, for all temperatures $\beta_e<\beta$. Note that the ferromagnetic solution $f_f$ has no entropy, $s_f=0$, which is here reflected by the fact that the potential $\phi_1$ depends upon $\beta_f$ and $x_f$ only through $m_f\equiv \beta_f x_f$. These observations allow us to focus on a simplified potential
\begin{equation}
\hat \phi(\beta_e,m)=\phi_1\left(\beta_e,x_e=\frac{m}{\beta_e},m_f=-m\right)
\end{equation}
which satisfies:
\begin{equation}
\partial_m\hat\phi=f_f-f_e,\qquad \partial_{\beta_e}\hat\phi=-ms_e.
\end{equation}

As with the BEC, the average error exponent is identified with the smallest value of $L_1$ such that $s_e\geq 0$ and $f_f -f_e\geq 0$. The present formulation is in fact equivalent to the presentation based on the replica method given in~\cite{SkantzosvanMourik03}. A remarkable consequence of the analysis is that the average error exponent is predicted to be the same for any $\beta\geq 1$. Indeed, both the glassy and the ferromagnetic free energies are temperature-independent for $\beta\geq \beta_g$. In particular, symbol and word-MAP are predicted to have same error exponents.

\begin{figure}
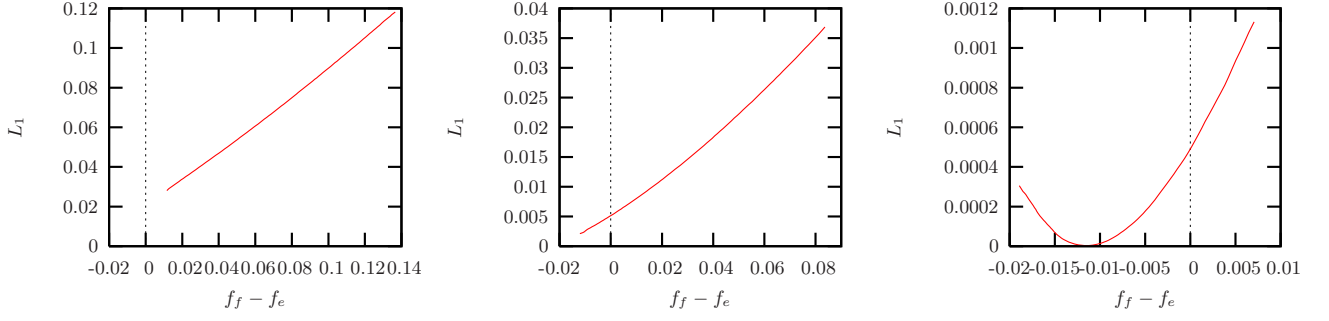

\begin{center}
\resizebox{.32\linewidth}{!}{\input{bscld1}}
\resizebox{.32\linewidth}{!}{\input{bscld2}}
\resizebox{.32\linewidth}{!}{\input{bscld3}}
\caption{\label{bsclargedev}\small Large deviation rate ${L_1}(f_f-f_e,s_e=0)$ as a function of the difference between the ferromagnetic and the non-ferromagnetic free energies, here for regular codes with $k=6$ and $\ell=3$ on the BSC. The thresholds are $p_{\rm 1RSB}\approx 0.058$ and $p_c\approx 0.100$. The three regimes are represented. From left to right: $p=0.045$, $p=0.07$ and $p=0.09$.}
\end{center}
\end{figure}

\begin{figure}
\begin{center}
\resizebox{.5\linewidth}{!}{\input{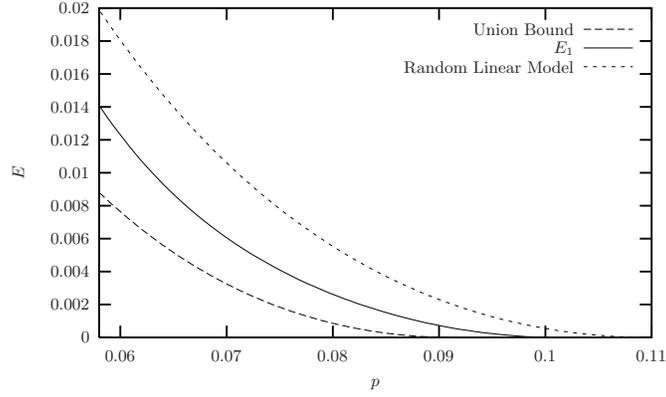}}
\caption{\label{bscerrexp}\small Average error exponent as a function of the noise level $p$ for the regular code ensemble with $k=6$ and $\ell=3$ through the BSC. Here $p_{\rm 1rsb}\approx 0.058$. The union bound~\eqref{unionbsc} and the random linear model ($k,l\to \infty$) limit~\eqref{RLMBSC} are also represented for comparison.}
\end{center}
\end{figure}

Based on the cavity equations given in Appendix~\ref{sec:BSCdetails}, the potential $\hat{\phi}$ can be computed numerically by population dynamics. As an illustration, we plot in Fig.~\ref{bsclargedev} the rate function $L_1(f_f-f_e, s_e=0)$ for a regular code with $k=6$, $\ell=3$. As in the case of BEC, three regimes can be distinguished, according to the value of $p$:
\begin{itemize}
\item $p<p_{\rm 1RSB}$: no zero-entropy RS solution typically exists, and $f_e<f_f$ for the metastable solutions.
\item $p_{\rm 1RSB}<p<p_d'$: no zero-entropy RS solution typically exists but the dominant metastable solutions have $f_e>f_f$.
\item $p_d'<p<p_c$: a zero-entropy RS solution is typically present.
\end{itemize}
The major difference with the BEC is that the threshold $p_d'$, defined by $p_d'=p_d(\beta_g(p_d'))$ does not coincide with the dynamical threshold $p_d(\beta)$: indeed here $p'_d$ is defined in relation to the existence of a solution with positive entropy, while, in the framework of BP, the dynamical arrest $p_d$ is related to the existence of a paramagnetic solution at decoding temperature $\beta^{-1}$~\cite{FranzLeone02}.
In Fig.~\ref{bscerrexp}, we plot the average error exponent for regular codes with $k=6$, $\ell=3$.

\subsection{The random code limit}
\subsubsection{Average error exponent}
As with the BEC, the  $k,\ell\vers\infty$ limit can be computed exactly, yielding
\begin{equation}\label{klinfinylimit}
E^{(1)}_1=L_1(f_f=f_e,s_e=0)=D(\delta_{GV}(R)\Vert p),
\end{equation}
where $\delta_{GV}(R)$ denotes the smallest solution to $R-1+H(\delta)=0$.
In this regime, errors are most likely to be caused by large noises driving the received message beyond the typical nearest-codeword distance.

As pointed out in~\cite{SkantzosvanMourik03}, a second ferromagnetic solution is present in this limit (see Appendix \ref{sec:BSCdetails} for details), yielding the error exponent:
\begin{equation}\label{eq:errexptypeI}
E^{(2)}_1=-\ln\frac{1}{2}\left(1+2\sqrt{p(1-p)}\right)-R\ln 2.
\end{equation}
Such a solution also exists for finite $k,\ell$, but is clearly unphysical (it predicts  negative exponents for $k=6$, $\ell=3$).
Yet it correctly describes the low $p$ phase \eqref{RLMBSC} in the $k,\ell\vers\infty$ limit, where failure is caused by the existence of one (or a few) unusually close codewords. In that sense it plays the same role as the energetic solution in the BEC analysis, with the difference that it is not extensible to any case with finite connectivities.
The critical noise $p_{e}$ below which such a scenario occurs is given by:
\begin{equation}\label{eq:pebsc}
\frac{\sqrt{p_e}}{\sqrt{p_e}+\sqrt{1-p_e}}=\delta_{GV}(R).
\end{equation}
We thus predict the average error exponent to be:
\begin{equation}
E_1(\rm RLM)=\left\{\begin{array}{ll}
D(\delta_{GV}(R)\Vert p) & \textrm{if }p<p_e<p_c,\\
-\ln\frac{1}{2}\left(1+2\sqrt{p(1-p)}\right)-R\ln 2 & \textrm{if }p<p_e.
\end{array}\right.
\end{equation}
This expression coincides with the exact result \eqref{RLMBSC} of the RLM.

\subsubsection{Typical error exponent}
The typical exponent of the RLM can be evaluated using the two-step potential:
\begin{equation}\label{eq:psibsc}
\e^{N\psi(\beta_e,m,y)}=\E_C\left[\e^{N y\hat\phi(\beta_e,m)}\right]=\int \ud \hat \phi\  e^{N(y\hat\phi-\L(\hat\phi,\beta_e,m))}.
\end{equation}
The details of the calculations by the cavity method are given in Appendix~\ref{sec:BSCdetails}.  As in the average case, two distinct solutions appear.
The first one is the counterpart of the solution discussed in section \ref{sec:avrgbsc}. It yields, in the random linear limit:
\begin{equation}
\psi(\beta_e,m,y)=y\hat\phi(\beta_e,m).
\end{equation}
A consequence of the linear dependence on $y$ is that $\hat\phi$ always takes the value obtained from the average calculation, irrespectively of $y$. Therefore, the average and typical error exponents coincide in this regime, and are given by \eqref{klinfinylimit}.

\begin{figure}
\begin{center}
\resizebox{.5\linewidth}{!}{\input{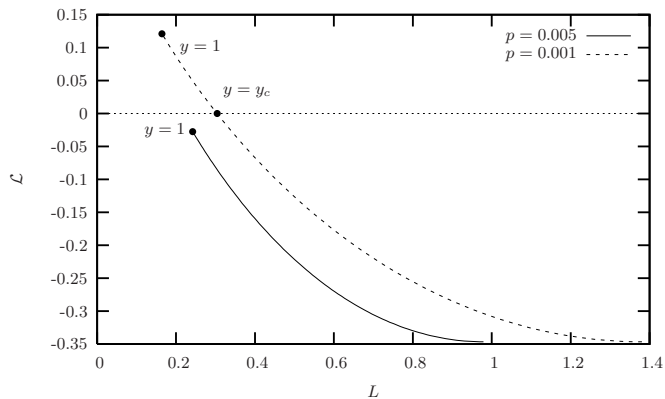}}
\caption{\label{fig:2stepbsc}\small Rate function $\L(L)$ for the RLM on the BSC with $R=1/2$ and $p=0.005>p_y$ (full curve), $p=0.001<p_y$ (dashed curve). }
\end{center}
\end{figure}

This solution is however only valid in the high noise regime ($p>p_e$). As in the average case, for low $p$, the errors in decoding are dominated by the presence of a sub-exponential (zero entropy) number of close codewords. The associated solution has for potential
\begin{equation}
\psi(y)=-yL-\L=(R-1)\ln 2+\ln\left[1+\left(2\sqrt{p(1-p)}\right)^y\right].
\end{equation}
We observe two types of behavior according to the value of $p$: for $p_y<p<p_e$, $\L(y)$ is negative for $0\leq y\leq 1$, whereas for $p<p_y$, it crosses $0$ at $y_c<1$ (see Fig.~\ref{fig:2stepbsc}). Interpreting, as in the BEC analysis (see section \ref{sec:typcav}), negative values of $\L$ as the evidence of a glassy transition in the space of codes, we deduce that the typical error exponent is given by $L(y_c)$ when $y_c<1$, in which case it differs from the average error exponent. To sum up:
\begin{equation}\label{eq:errexptypeIb}
E_0({\rm RLM})=\left\{\begin{array}{ll}
L(y_c)=-\delta_{GV}(R)\ln\left[2\sqrt{p(1-p)}\right]&\textrm{if }p<p_y,\\
L(y=1)=E_1({\rm RLM})&\textrm{if }p_y<p<p_c,\end{array}\right.
\end{equation}
where the critical noise $p_y(R)$ is solution of:
\begin{equation}\label{eq:pybsc}
\frac{2\sqrt{p_y(1-p_y)}}{1+2\sqrt{p_y(1-p_y)}}=\delta_{GV}(R).
\end{equation}
This exponent coincides with the RLM limit of the union bound \eqref{eq:ubkilibsc}, and is rigorously established \cite{BargForney02} to be the correct typical error exponent on the BSC.

\section{Conclusion}

Since Shannon laid the basis for information theory, the analysis of 
error-correcting codes has been a major subject of study in this 
field of science~\cite{Gallager62}. Error-correcting 
codes aim at reconstructing signals altered by noise. Their performance is measured by their error probability, i.e. the probability that they fail in accomplishing this task. For block codes, 
where the messages are taken from a set of $2^M$ codewords of length 
$N$, it is known that when the rate $R=M/N$ is below the channel 
capacity $R_c$, the probability of error behaves, in the limit of large 
$N$, at best, as $P_e\sim \exp(-NE(R))$~\cite{Gallager62}. This error exponent 
$E(R)$, also called reliability function, provides a particularly concise characterization of performance.

For a given code ensemble, two 
classes of error exponents can generally be distinguished, due to the presence of two levels of 
``disorder'', one associated with the choice of the code itself, and a second associated with the realization of the noise. {\it Average error 
exponents} correspond to take the error probability $P_e$ with respect 
to these two levels simultaneously, while {\it typical error 
exponents} refer to fixed, typical, codes.

In the present paper, we tackled the computation of these two error 
exponents for a particular class of block codes, the low-density 
parity-check (LDPC) codes, with two particular channels, the binary 
erasure channel (BEC) and the binary symmetric channel (BSC). 
We considered decoding under maximum-likelihood decoding, the best conceivable decoding procedure. We framed the problem in terms of large deviations, and applied a recently proposed extension of the cavity method designed to probe atypical events in systems defined on  random graphs~\cite{Rivoire05}. This method provides an alternative to the replica method used in~\cite{SkantzosvanMourik03} to address similar problems, with the advantage of being based on explicitly formulated probabilistic assumptions. With respect to this earlier contribution, our work offers several clarifications, notably on the nature of the different phases, and various extensions, notably to the BEC channel. With this particular channel, our results are analytical, and, in the high-noise regime, we conjecture them to be exact. Recent mathematical results on the typical phase diagram \cite{MeassonMontanari04} foster hope for a confirmation of our results in that context.

From a statistical physics perspective, error exponents are interesting for the richness of their phase diagram, which comprises two phase transitions of different natures. These transitions are observed when the level of noise $p$ is  varied at fixed rate $R$ (or, equivalently in the special case of random codes, when the rate $R$ is varied at fixed $p$). Close to the static threshold, for $p_e<p<p_c$, errors are mostly due to the proliferation of many incorrect codewords in the  vicinity of the received message. We interpreted this feature in terms  of the presence of a glassy phase, and, accordingly, we were able to describe this regime by considering a one-step replica symmetry breaking (1RSB) approach. Below $p_e$, errors become dominated by the effect of single isolated codewords, which we attributed to a transition towards a ferromagnetic state, or 1RSB to RS transition. The noise $p_e$ has its counterpart in the ``critical rate'' $R_e$ of information theory~\cite{Gallager62}, which marks the point below which only bounds on the reliability function are known. The replica symmetric (RS) approach we employed to investigate the regime $p<p_e$ also turns out to be only approximate, except in the limit of infinite connectivity, where we recovered the error exponents of random linear codes~\cite{BargForney02}. We also described a second transition occurring at $p_y<p_e$, below which atypical codes come to dominate the average exponent, causing it to differ from the typical error exponent. As it takes place in the space of graphs, this is an example of critical phenomenon whose description is not accessible to the standard cavity method~\cite{MezardParisi01}, but only to its extension to large deviations~\cite{Rivoire05} (see also~\cite{RivoireBarre06} for an other example). However, this second transition should be taken with utmost care, as it relies on an approximate ansatz.

The numerous efforts made in the information theory community to account 
for the low rate regime $R<R_e$ have so far resulted only in upper and 
lower bounds for the reliability function~\cite{Berlekamp02}. Maybe not too surprisingly, this is also the region of the phase diagram where our methods encounter difficulties. Several examples are however now available which demonstrate that statistical  physics methods can provide exact solutions to notoriously difficult mathematical problems. The solutions thus obtained generally sharpen 
our comprehension both of the system at hand and of the 
techniques themselves, besides often paving the way for rigorous 
derivations. In the light of some recent such achievements, extending the present 
statistical physics approach to reach a thorough understanding of error 
exponents seems to us a valuable challenge.

\acknowledgments

The work of T.M. was supported in part by the EC through the network
MTR 2002-00319 `STIPCO' and the FP6 IST consortium `EVERGROW'. O.R. is a fellow
of the Human Frontier Science Program.

\appendix

\section{A note on the exponential scaling}\label{sec:scaling}

The thermodynamic approach is based on the assumption that the
leading contribution to the probability of error decays
exponentially with $N$. However, as initially shown by
Gallager, for ensembles of LDPC codes, the probability of error decays only polynomially in $N$ to the leading order. In physical terms, this is due to a few codes
(whose number is a polynomial in $N$) which display a second, metastable, ferromagnetic state at a smaller distance
from the ground state state (corresponding to the correct
codeword) than the numerous configurations forming the
paramagnetic state.

To overpass this spurious effect in the simplest, yet purely
theoretical way, Gallager focused on the so-called ``expurgated
ensemble'' where the half of the codes with smallest minimum
distance is disregarded. On this restricted ensemble which
excludes the codes with multiple ferromagnetic states, the error
probability decays now exponentially in $N$ at the leading
order and can be characterized with an average error exponent.
Needless to say, this construction only makes sense as a convenient theoretical way to access
good codes.

As the large deviation method automatically overlooks any polynomial
contribution, its results actually apply to the ``expurgated
ensemble''. This is however only true to the extend that
the expurgation does not affect the distribution of graphs in the
ensemble (i.e., does not change the distribution of degrees, of
loops, etc.). This is presumably the case, as supported by the construction presented in~\cite{MourikKabashima03}, where an expurgated
ensemble much tighter than Gallager's one is defined by explicitly
associating to any random code an expurgated code obtained by
modifying only a number $O(1)$ of small loops.

\section{Random Linear Model}\label{sec:RLM}

\subsection{Definition}

A parity-check code is defined by a $M\times N$ matrix $A$ over $\Z_2$ and its codewords are the vectors $\bx=(x_1,\dots,x_N)$ satisfying $A\bx=0$. Code ensembles are therefore subsets of the set of all $2^{MN}$ possible matrices. Taking this complete set (with all possible matrices having same probability) defines the so-called {\it random linear model} (RLM). In contrast with LDPC codes, since a typical matrix from the RLM is not sparse, the belief propagation algorithm cannot be used to decode. While of little practical interest due to this absence of efficient decoding algorithm, the RLM has however two major theoretical advantages, both originating from its ``maximally random'' nature: typical codes from the RLM saturate the Shannon bounds, and error exponents can be derived rigorously. We review here some of the established results, which we used in the main text as a reference point to compare our non-rigorous results. Error exponents for the RLM are indeed expected to provide upper bounds for error exponents of LDPC ensemble, which are reached only in the limit of infinite connectivity $k,l\to\infty$ (this limit is similar to that in which $p$-spin models approach the random energy model when $p\to\infty$~\cite{Montanari01}).

\subsection{Weight enumerator function}

We first characterize the geometry of the space of codewords by means of the so-called {\it weight enumerator function}. Given a code $C$ with matrix $A$, this function gives the number $\N_C(d)$ of codewords $\bx$ at (Hamming) distance $d=|\bx|\equiv\sum_{i=1}^N x_i$ from the origin:
\begin{equation}\label{wef}
\N_C(d)=\sum_{\bx} \delta\left(d,\sum_{i=1}^N x_i\right) \delta(A\bx,0),
\end{equation}
where the sum is over all codewords, and $\delta(x,y)$ enforces the constraint $x=y$. The {\it average} weight enumerator function is obtained by averaging over the code ensemble and satisfies
\begin{equation}
\overline{\N}(d)\equiv \E_C[\N_C(d)]=\binom{N}{d}2^{-M}\asymp \e^{N\Sigma(R,\delta=d/N)},\qquad \Sigma(R,\delta)=(R-1)\ln 2+H(\delta),
\end{equation}
where the limit of infinite block-length $N\vers\infty$ is taken with $M=N(1-R)$ and $d=Nx$. The exponent $\Sigma(R,x)$ defines the so-called {\it average weight enumerator exponent}. A critical distance is the distance $\delta_{GV}(R)$ defined as the smallest $\delta>0$ such that $\Sigma(R,\delta)=0$. Codewords at distance $d=N\delta$ with $\delta>\delta_{GV}(R)$ proliferate exponentially. On the other hand, the probability of existence of a codeword at distance $d=N\delta$ with $\delta<\delta_{GV}(R)$ is upper-bounded by $\overline{\N}(d)$, and thus decays exponentially with $N$. Consequently, for any $\epsilon(N)$ such that $\epsilon(N)\vers\infty$ (e.g. $\epsilon(N)=\sqrt{N}$), only an exponentially small fraction of the codes in the ensemble have a minimal non-zero distance $d=N\delta$ smaller than $N\delta_{GV}(R)-\epsilon(N)$. Excluding these ``worst'' codes from the RLM defines the {\it expurgated RLM ensemble}.

\subsection{Average error exponent over the BEC}

Due to the group symmetry of the set of codewords, we can assume without loss of generality that the transmitted codeword is $(0,\ldots,0)$. For a given realization of the disorder due to a BEC, we denote by $E\subset \{1,\ldots,N\}$ the subset of erased bits in the received string, and $d$ the number of elements in $E$. If $A$ is the $M\times N$ matrix representing the code, the sub-matrix $\tilde A^E$ induced by $A$ on $E$ defines the decoding-CSP problem: decoding is impossible if and only if the kernel of $\tilde A^E$ is non-zero. When all matrices $A$ are sampled with uniform probabilities as in the RLM, the sub-matrices $\tilde A^E$ are also represented with uniform probability. Given a noise realization $E$ of magnitude $d$, the error probability is the probability that a random $M\times d$ matrix $\tilde A^E$ is non-injective,
\begin{equation}
\E_C[\P^{(B)}_N(\mathbf{0})]=\sum_{d=0}^{N}\binom{N}{d}p^{d}(1-p)^{N-d}\P(\exists \bx\neq \mathbf{0}\textrm{ such that } \tilde A^E \bx=\mathbf{0}).
\end{equation}
When $d>M$, $\tilde A^E$ is necessarily non injective. When $d\leq M$ on the other hand, a straightforward inductive argument~\cite{DiProietti02} gives
\begin{equation}\label{pm1}
\P(\exists \bx\neq \mathbf{0}\textrm{ such that } \tilde A^E \bx=\mathbf{0})=1-\prod_{i=0}^{d-1}(1-2^{i-M})
\end{equation}
consequently, the {\it exact} expression for the average error probability of the RLM reads
\begin{equation}
\E_C[\P^{(B)}_N(\mathbf{0})]=\sum_{d=0}^{M}\binom{N}{d}p^{d}(1-p)^{N-d}\left(1-\prod_{i=0}^{d-1}(1-2^{i-M})\right)+\sum_{d=M+1}^{N}\binom{N}{d}p^{d}(1-p)^{N-d}.
\end{equation}
In the  $N\vers\infty$, this expression can be evaluated by the saddle point method. When $p<(1-R)/(1+R)$, the dominant contribution comes from the first sum, with
\begin{equation}
\sum_{d=0}^{M}\binom{N}{d}p^{d}(1-p)^{N-d}\left(1-\prod_{i=0}^{d-1}(1-2^{i-M})\right)\asymp e^{-N[(1-R)\ln 2-\ln(1+p)]},
\end{equation}
and the typical number of errors $d=N 2p/(1+p)$. When $p>(1-R)/(1+R)$, (and $p<1-R$ to stay below the capacity), the dominant contribution comes from the second sum, with
\begin{equation}
\sum_{d=M+1}^{N}\binom{N}{d}p^{d}(1-p)^{N-d}\asymp e^{-ND(1-R||p)}.
\end{equation}
and the typical number of errors $d=N(1-R)$. We thus obtain for the average error exponent of the RLM the expression given in Eq.~\eqref{RLM},
\begin{equation}\label{RLM2}
E_{1}(\textrm{RLM})=\left\{\begin{array}{ll}(1-R)\ln 2-\ln(1+p) 
& \textrm{if }p<\frac{1-R}{1+R},\\
D(1-R||p)&\textrm{if }\frac{1-R}{1+R}<p<1-R.\end{array}\right.
\end{equation}
In physical terms, the transition between the two regimes can be interpreted as a transition between a ferromagnetic (RS) phase and a glassy (1RSB) phase. In the large noise regime, $p>(1-R)/(1+R)$, the error is indeed most probably due to the noise driving the received string into a ``glassy phase'' of exponentially numerous incorrect codewords, as reflected by the fact that then $\P(\exists \bx\neq \mathbf{0}\textrm{ such that } \tilde A^E \bx=\mathbf{0})=1$. In contrast, in the low noise regime, $p<(1-R)/(1+R)$, the error is most probably due to the noise driving the received string into a ``ferromagnetic phase'' where an isolated incorrect codeword happens to be closer than the correct codeword; this is reflected by the fact that $\P(\exists \bx\neq \mathbf{0}\textrm{ such that } \tilde A^E \bx=\mathbf{0})$ differs from 1 only by an exponentially small term in $N$, as seen from Eq.~\eqref{pm1}.

\subsection{Average error exponent over the BSC}

With the binary symmetric channel (BSC), starting again from the transmitted codeword is $(0,\ldots,0)$, the received string $\by$ cannot be decoded if there exists $\bx\neq\mathbf{0}$ such that $A\bx=\mathbf{0}$ and $|\bx-\by|<|\by|$. Denoting $P_e(\by)$ the probability of this event, the probability of error is
\begin{equation}
\E_C[\P^{(B)}_N(\mathbf{0})]=\sum_{d=0}^{N}\binom{N}{d}p^d(1-p)^{N-d}P_e(\by^{(d)}),
\end{equation}
where $\by^{(d)}$ is a generic string of weight $d$, e.g. $y_i=1$ if $i\leq d$, $y_i=0$ if $i>d$. If $d/N>\delta_{GV}(R)$, $P_e(\by^{(d)})$ goes to one in the infinite block-length limit. Although no published proof is available in the literature, it is reported as proved~\cite{BargForney02} that, when $d/N<\delta_{GV}(R)$, $P_e(\by^{d})$ is asymptotically equivalent to its union bound approximation (see the following appendix), i.e.,
\begin{eqnarray}
P_e(\by^{(d)})&\sim& \E_C\left[\sum_{\bx\neq\mathbf{0}}\theta(d-|\bx-\by^{(d)}|)\delta(A\bx,\mathbf{0})\right]\\
&\sim&  \sum_{i=0}^d \E_C[\mathcal{N}_C(i,\by^{(d)})]\\
&\sim& \E_C[\mathcal{N}_C(d,\by^{(d)})]
\end{eqnarray}
where $\mathcal{N}_C(i,\by^{(d)})$ is the number of codewords at distance $i$ from $\by^{(d)}$, and $\theta(x)=1$ if $x>0$, and 0 otherwise. Straightforward combinatorics shows that the asymptotic behavior of $\E_C\mathcal{N}_C(i,\by^d)$ is given by the standard weight enumerator exponent $\Sigma(R,i/N)$. In the limit $N\vers\infty$ where $\delta=d/N$ is kept fixed, a saddle-point evaluation leads to the following expression of the average error exponent:
\begin{eqnarray}
E_1(\textrm{RLM})&=&-\max_{\delta<\delta_{GV}}\left[\Sigma(R,\delta)-D(\delta||p)\right]\\
&=&\left\{\begin{array}{ll}(1-R)\ln 2 -\ln\left[1+2\sqrt{p(1-p)}\right] &\textrm{ if }\frac{\sqrt{p}}{\sqrt{p}+\sqrt{1-p}}<\delta_{GV}(R),\\ D(\delta_{GV}(R)||p) & \textrm{otherwise.}\end{array}\right.\label{RLMBSC}
\end{eqnarray}
This results with two distinct regime is very similar to that obtained previously for the BEC.

\section{Union bounds}\label{sec:union}

The so-called {\it union bound exponent} is a rigorous lower bound of the average error exponent in the expurgated ensemble. We show in this appendix how the average weight enumerator exponent of (regular) LDPC codes can be used to derive this union bound exponent, for both the BEC and the BSC. We will thus recover results first established by Gallager in \cite{Gallager62,Gallager68}. In a nutshell, the idea of the union-bound is to upper-bound the probability that at least one (bad) codeword causes an error by the sum of the probabilities that each does. Remarkably, this union bound turns out to be tight for the RLM ensemble.

\subsection{Weight enumerator function}

The weight enumerator function (see Eq.~\eqref{wef} for the definition) of regular LDPC codes with $k=6$ and $\ell=3$ was computed in \cite{Gallager62} and reads:
\begin{eqnarray}
\E_C[\mathcal{N}_C(d)]&=&\sum_{\bx} \delta(|\bx|,d)\E_C\left[\delta(A\bx=\mathbf{0})\right]=\binom{N}{d}\E_C\left[\delta(A\bx^{(d)}=\mathbf{0})\right]\\
\E_C[\mathcal{N}_C(d=\delta N)]&\asymp& \e^{N\Sigma(k,l,\delta)},\\
\textrm{with } 
\Sigma(k,l,\delta)&=&\min_\mu \left(2\mu\ell \delta+(1-\ell)H(\delta)+\frac{\ell}{k}\ln C(\mu)\right),\\
\textrm{and }
C(\mu)&=&\frac{1}{2}\left[{\left(1+\e^{-2\mu}\right)}^{k}+{\left(1-\e^{-2\mu}\right)}^{k}\right].
\end{eqnarray}
We introduce $\delta_m$, the smallest $\delta$ such that $\Sigma(k,l,\delta)\geq 0$.
By construction, the average enumerator exponent in the expurgated ensemble is 
\begin{equation}
\Sigma_{\textrm{exp}}(k,l,\delta)=\left\{\begin{array}{ll}\Sigma(k,l,\delta)&\textrm{if }\Sigma(k,l,\delta)>0 (\textrm{\ i.e. if }\delta>\delta_m),\\-\infty&\textrm{otherwise.}\end{array}\right.
\end{equation}
This expurgated {\it average} enumerator exponent $\Sigma_{\textrm{exp}}(k,l,\delta)$ is believed to coincide with the {\it typical} enumerator exponent \cite{Condamin02,DiMontanari04}.

\subsection{Union bound for the BEC}

Given the set $E$ of erased bits, we want to estimate the probability $P_e(d)$ that the CSP-decoding problem has at least two solutions, when a code $C$ is drawn at random from its ensemble. We call $A$ the matrix characterizing $C$, $\tilde A^E$ the sub-matrix induced by $A$ on $E$, and $d$ the number of erased bits. The union bound consists in the following inequality:
\begin{eqnarray}
P_e(d)&=&\P(\exists \tilde\bx\in\{0,1\}^d\neq\mathbf{0}\textrm{ such that }\tilde A^E \tilde\bx=\mathbf{0})\label{eqP}\\
&\leq& \min\left[\sum_{\tilde\bx\neq\mathbf{0}}\P(\tilde A^E \tilde\bx=\mathbf{0}),1\right]
\end{eqnarray}
Let $w=|\tilde \bx|$ and $\bx$ be constructed from $\tilde \bx$ by setting $x_i=\tilde x_i$ for $i\in E$, $x_i=0$ otherwise: $\tilde \bx$ belongs to the kernel of $\tilde A$ if and only if $\bx$ belongs to the kernel of $A$. The probability of the latter event reads
\begin{equation}
\E_C[\mathcal{N}_C(w)]\binom{N}{w}^{-1}.
\end{equation}
The error probability is consequently bounded by:
\begin{eqnarray}
\E_C[\P_N^{(B)}]&=&\sum_{d=0}^{N}\binom{N}{d}p^d(1-p)^{N-d}P_e(d)\label{eqP2}\\
&\leq &\sum_{d=0}^N\binom{N}{d}p^d(1-p)^{N-d}\min\left[\sum_{w=0}^d \binom{d}{w}\E_C[\mathcal{N}_C(w)]\binom{N}{w}^{-1},1\right].\label{UB}
\end{eqnarray}
In the infinite block-length limit, a saddle-point estimate yields, as upper-bound for the expurgated average error exponent, the exponent
\begin{equation}\label{unionbec}
\begin{split}
E_{\textrm{exp}}(k,l)\geq E_{UB}&=-\max_{\delta} \left\{-D(\delta\Vert p)+\min\left[\max_{\omega} \left(\Sigma(\omega)+H\left(\frac{\omega}{\epsilon}\right)-H(\omega)\right),0\right]\right\}\\
&=-\max_{\delta<\delta_{UB}} \left\{-D(\delta\Vert p)+\max_{\omega>\delta_m}\min_\mu \left[H\left(\frac{\omega}{\delta}\right)+2\mu\ell \omega-\ell H(\omega)+\frac{\ell}{k}\ln C(\mu)\right]\right\}
\end{split}
\end{equation}
where $\delta=d/N$, $\omega=w/N$, and $\delta_{UB}$ is the largest $\delta$ such that $\max_\omega \left(\Sigma(\omega)+H\left(\frac{\omega}{\delta}\right)-H(\omega)\right)$ is non-positive.

\begin{figure}
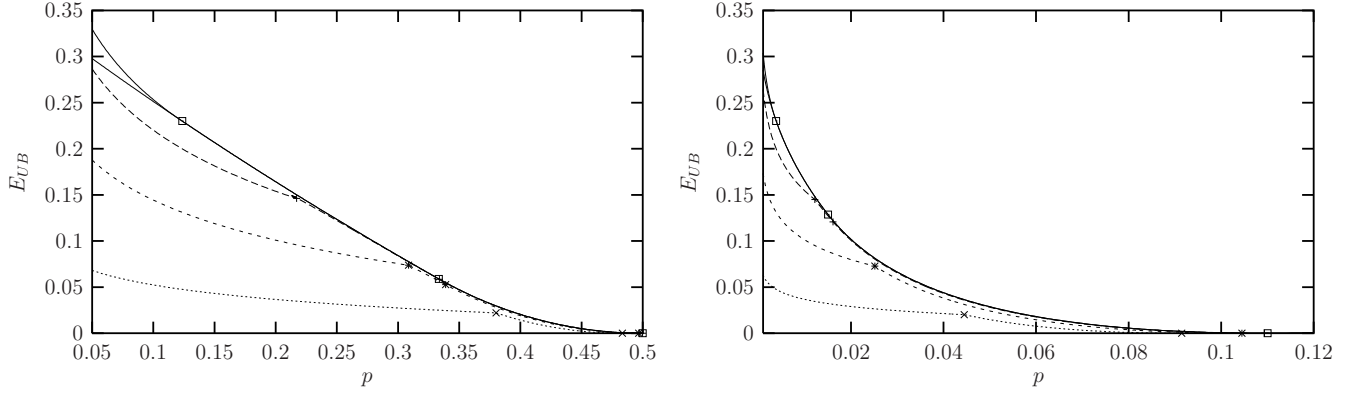

\centering
\large
\resizebox{.49\linewidth}{!}{\input{ubbec}}
\resizebox{.49\linewidth}{!}{\input{ubbsc}}
\caption{\label{fig:union}\small Expurgated union bounds for the BEC (left) and the BSC (right). From bottom to top, $(k,\ell)=(6,3),(8,4),(12,6)$ and the RLM limit, expurgated (top full curve) and not expurgated (bottom full curve) with $R=1/2$. The points indicate the transition between the three regimes, as well as $e_{UB}$.}
\end{figure}

As $p$ is varied, three regimes can be distinguished . For small $p$, the maximum over $\omega$ is reached on the boundary $\delta_m$, meaning that errors are dominated by the nearest codewords. For large $p$ instead, the maximum over $\delta$ is reached at $\delta_{UB}$, in which case the union bound is simply replaced by $1$, physically corresponding to a large number of bad codewords arising from the large amplitude of the noise. Finally, in the intermediate region of $p$, the extremum is reached in the interior of the $(\delta,\omega)$ domain. Note that this last regime is not always present when $k$ and $\ell$ are too small (for $k=6$ and $\ell=3$ in particular). These three regimes are given in the limit $k,\ell\vers \infty$ by:
\begin{equation}\label{eq:ubkilibec}
E_{0}(\textrm{RLM})=\left\{\begin{array}{ll}
-\delta_{GV}(R)\ln p& \textrm{if }p<p_y \\
(1-R)\ln 2-\ln(1+p) 
& \textrm{if }p_y<p<\frac{1-R}{1+R},\\
D(1-R||p)&\textrm{if }\frac{1-R}{1+R}<p<1-R.\end{array}\right.
\end{equation}
with $p_y$ defined as in \eqref{eq:py}.
Union bounds for the BEC are plotted in Fig.~\ref{fig:union} for several regular ensembles.

\subsection{Union bound for the BSC}

The union bound for the BSC is derived following the same steps than for the BEC. The counterpart of Eq.~\eqref{eqP} reads
\begin{equation}
P_e(d)=\P(\exists{\bx\neq 0}\textrm{ such that }|\bx-\by^{(d)}|<d \textrm{ and }A \bx=0),
\end{equation}
where $\by^{(d)}$ is a generic string of weight $d$. Let $\bx$ be a string a weight $w$ and $Q(w,d,g)$ be the probability for $\by^{(d)}$ to be at distance $g$ from $\bx$, conditioned on $|\by^{(d)}|=d$:
\begin{equation}
Q(w,d,g)=\binom{w}{(d-g+w)/2}\binom{N-w}{(d+g-w)/2}{\binom{N}{d}}^{-1}.
\end{equation}
The probability for $\by^{(d)}$ to be at distance $g$ from any codeword $\bx$ is upper-bounded by
\begin{equation}
\sum_w \E_C[\N_C(w)] Q(w,d,g)
\end{equation}
and we can write
\begin{equation}
P_e(d)\leq \min\left[\sum_{w,g} \E_C[\N_C(w)] Q_C(w,d,g),1\right]\asymp \min\left[\sum_{w} \E_C[\N_C(w)] Q_C(w,d,d),1\right].
\end{equation}
From this inequality and Eq.~\eqref{eqP2}, we obtain the union bound for the error exponent via the saddle-point method:
\begin{equation}
\begin{split}
E_{\textrm{exp}}(k,l)\geq E_{UB}&=-\max_{\delta} \left\{-D(\delta\Vert p)+\min\left[\max_{\omega} \left(\Sigma(\omega)+L(\omega,\delta,\delta)\right),0\right]\right\}\\
&=-\max_{\delta<\delta_{UB}}\left\{-D(\delta \Arrowvert p)+ \max_{\omega >\delta_m}\min_\mu\left[2\mu\ell \omega+(1-\ell)H(\omega)+\frac{\ell}{k}\ln C(\mu)+L(\omega,\delta,\delta)\right]\right\}\label{unionbsc},\\
L(\omega,\delta,\gamma)&=\omega H\left(\frac{\delta-\gamma+\omega}{2\omega}\right)+(1-\omega)H\left(\frac{\delta+\gamma-\omega}{2(1-\omega)}\right)-H(\delta).
\end{split}
\end{equation}
As for the BEC, three regimes can be distinguished, according to the value of $p$. In the limit $k,\ell\vers\infty$, these three regimes are:
\begin{equation}\label{eq:ubkilibsc}
E_0(\textrm{RLM})=\left\{\begin{array}{ll}
-\delta_{GV}(R)\ln\left[2\sqrt{p(1-p)}\right]&\textrm{ if }p<p_y,\\
(1-R)\ln 2 -\ln\left[1+2\sqrt{p(1-p)}\right] &\textrm{ if }p_y<p<p_e,\\
D(\delta_{GV}(R)||p) & \textrm{ if }p_e<p<\delta_{GV}(R)\end{array}\right.
\end{equation}
where $p_y$ and $p_e$ are given by \eqref{eq:pybsc} and \eqref{eq:pebsc}.

Union bounds for the BSC are plotted in Fig.~\ref{fig:union}.

\section{Irregular codes}\label{sec:irrdetails}

\subsection{Definition of the ensemble}
In this appendix we discuss the generalization to irregular graphs. We shall only treat the entropic large deviations with the BEC, but our arguments can easily be generalized to the other cases. With irregular codes, it is necessary to specify more precisely the definition of the ensemble. The usual definition is via the degree distributions $v_\ell$ and $c_k$. It is however possible to define different ensembles having same distribution and sharing the same typical properties, but differing at the level of atypical 
properties, including error exponents (see also ~\cite{Rivoire05} for similar non-equivalences in an other context).   

The simplest construction takes all factor graphs with exactly $v_\ell
N$ checks of degree $\ell$, $c_k M$ variables of degree $k$, and
pick them with uniform probability. Such ensembles are
used to build actual codes, and we shall therefore
analyze them with some details. 

\subsection{Average error exponent}
We revisit the arguments of section \ref{avrgbec} and emphasize the differences with the regular case.

A crucial modification is the introduction of Lagrange multipliers enforcing the number of nodes of each degree. Call $N_\ell$ the number of variables of degree $\ell$, and $M_k$ the number of checks of degree $\ell$. Denote $n_\ell=N_\ell/N$, $m_k=M_k/N$. The rate $L_1$ is now a function of the $n_\ell$ and $m_k$. Its multiple Legendre transform is defined as:
\begin{equation}
\begin{split}
&\phi(x,\{\lambda_\ell\},\{\nu_k\})\doteq xs+\sum_{\ell}\lambda_{\ell}n_{\ell}+\sum_k \nu_k m_k-L_1\\
\textrm{with}\quad &x=\partial_s L_1\qquad \lambda_\ell=\partial_{n_\ell}L_1\qquad \nu_k=\partial_{m_k}L_1
\end{split}
\end{equation}

Let us consider the addition of a new bit. $\ell$ checks are added along with it, where $\ell$ is drawn with probability $v_\ell$. Each of these checks, in turn, is connected to $k_a-1$ old bits ($a=1,\ldots,\ell$), where $k_a$ is drawn with probability $k_ac_{k_a}/\langle k\rangle$. Eq.~\eqref{phi} is modified in the following way:
\begin{equation}\label{phiirr}
\begin{split}
\phi(x,\{\lambda_\ell\},\{\nu_k\})&= \ln\sum_{\ell}v_{\ell}\sum_{\{k_1,\ldots,k_\ell\}}\prod_{a=1}^{\ell}\frac{k_ac_{k_a}}{\langle k\rangle}\int \ud \Delta S\, P^{(\ell,k_1,\ldots,k_\ell)}_{\circ+\square\in\circ}(\Delta S)\exp\left[x\Delta S+\sum_{a=1}^\ell((k_a-1)z_{k_a}+\nu_{k_a})+\lambda_\ell\right]\\
z_k&=-\frac{1}{k}\ln\int \ud\Delta S\,P^{(k)}_\square(\Delta S)\, e^{x\Delta S+\nu_k}
\end{split}
\end{equation}

The addition of a variable of degree $\ell$ is reflected by a factor $\e^{\lambda_\ell}$, and the addition of a check of degree $k$ by a factor $\e^{\mu_k}$.
Call $k$-{degree} the degree of a variable with respect to checks of degree $k$. Here $z_k$ is related to the increase of $k$-degrees in the ensemble.
Let us consider for a moment a more general setting, where the ensemble is determined by the $k$-degree distributions, denoted by $v^{(k)}_\ell$~\cite{footnoteInourcase}. Then $z_k$ is defined by
\begin{equation}
z_k=\sum_{\ell}\delta v^{(k)}_\ell \frac{\partial L_1(s,\{v^{(k)}_{\ell}\})}{\partial v^{(k)}_{\ell}}
\end{equation}
where $\delta v^{(k)}_\ell=v^{(k)}_{\ell-1}-v^{(k)}_\ell$. $z_k$ is obtained in a very similar way as $z$ in \eqref{calcz}:
\begin{equation}
z_k=-\frac{1}{k}\ln\int \ud\Delta S\,P^{(k)}_\square(\Delta S)\, e^{x\Delta S},
\end{equation}
where $P^{(k)}_\square(\Delta S)$ now depends on the degree $k$.

The cavity equation \eqref{cavity} is modified in a very similar way as the expression of $\phi_1$ in \eqref{phiirr}.
The inversion of the Legendre transformation allows to recover the relevant quantities:
\begin{equation}\label{quanttight}
s=\partial_x\phi\qquad n_\ell=\partial_{\lambda_\ell}\phi\qquad m_k=\partial_{\nu_k}\phi
\end{equation}

Replacing $P^{(\ell,k_1,\ldots,k_\ell)}_{\circ+\square\in\circ}(\Delta S)$ and $P^{(k)}_\square(\Delta S)$ by their values, we obtain:

\begin{equation}\label{eq:becirr}
\begin{split}
&\phi_1=xs-L_1=\ln\left[v(A)+p(2^x-1)v(B)\right]\\
\textrm{with }\quad &A=\e^{\lambda_\ell}\sum_k\frac{kc_k}{\bar k}\e^{(k-1)z_k+\nu_k}\left[2^{-x}+(1-2^{-x})(1-\nu)^k\right],\quad
B=2^{-x}\e^{\lambda_\ell}\sum_{k}\frac{kc_k}{\bar k}\e^{(k-1)z_k+\nu_k}(1-(1-\nu)^{k-1}),\\
&z_k=-\frac{1}{k}\ln\left[2^{-x}+(1-2^{-x})(1-\nu)^k\right]-\frac{\nu_k}{k},\quad \nu=\frac{p2^xv'(B)}{v'(A)+p(2^x-1)v'(B)}.
\end{split}
\end{equation}

To evaluate $L_1$ as a function of $s$, we simply need to tune the parameters $\lambda_\ell$ and $m_k$ such that the conditions $n_\ell=v_\ell$ and $m_k=\alpha c_k$ are satisfied. 

\begin{figure}
\begin{center}
\resizebox{.5\linewidth}{!}{\input{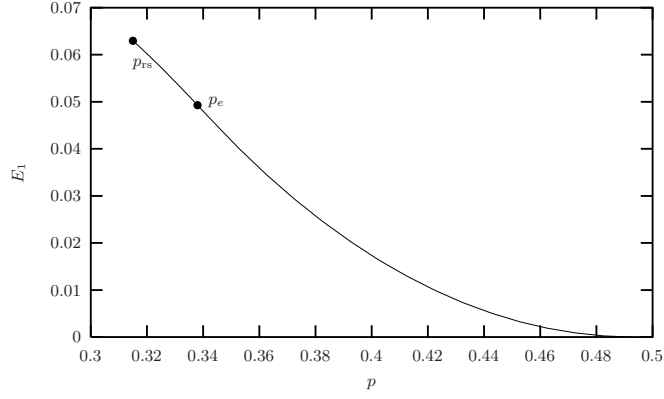}}
\caption{\label{fig:irregular}\small
Average error exponent of a given code as a
function of the noise level $p$ for irregular codes with
$c_k=(1/2)(\delta_{k,6}+\delta_{k,8})$ and
$v_\ell=(1/2)(\delta_{\ell,3}+\delta_{k,4})$ through the BEC.
}
\end{center}
\end{figure}

In Fig.~\ref{fig:irregular}, we represent the error exponent for the irregular ensemble with $v(x)=(1/2)x^3+(1/2)x^4$,
$c(x)=(1/2)x^6+(1/2)x^8$.

\section{Calculations in the BSC}\label{sec:BSCdetails}

\subsection{Belief Propagation and the Bethe approximation}
In this section we write down the BP equations for a given code over the BSC, or equivalently the cavity equations at the RS level. The expression of the free energy is also given.

The cavity equations read:
\begin{equation}\label{bpbsc}
\begin{split}
p_{\t_i}^{(i\to a)}&\propto\prod_{b\in i-a}q_{\t_i}^{(b\to i)}e^{-\beta h_i\t_i},\\
q_{\t_i}^{(b\to i)}&=\sum_{\t_{b-i}}\prod_{j\in b-i}p_{\t_j}^{(j\to b)}\delta[\t_b=1]
\end{split}
\end{equation}
$p_{\t_i}^{(i\to a)}$ is the probability that the variable $i$ takes the value $\t_i$ in the absence of $a$, and $q_{\t_i}^{(b\to i)}$ is proportional to the probability that the variable $i$ takes the value $\t_i$ when connected to $b$ only.

Denoting $p_{\t_i}^{(i\to a)}=\e^{\beta h_{i\to a}\sigma_i}/\cosh \beta h_{i\to a}$ and $q_{\t_i}^{(b\to i)}=\e^{\beta u_{b\to i}\t_i}/\cosh \beta u_{b\to i}$, the cavity equations simplify to:
\begin{equation}
\begin{split}
h_{i\to a}&=\hat h(h_i,\{u_{b\vers i}\})\equiv h_i+\sum_{b\in i-a}u_{b\to i}\\
u_{b\to i}&=\hat u(\{h_{j\vers b}\})\equiv \frac{1}{\beta}\mathrm{atanh}\left(\prod_{j\in b-i}\tanh\beta h_{j\vers b}\right)
\end{split}
\end{equation}

The local magnetization is given by $\langle \s_i\rangle=\tanh \beta H_i$, with $H_i=h_i+\sum_{a\in i}u_{a\to i}$. The Bethe approximation to the free energy reads:
\begin{equation}
F_{\textrm{RS}}(\beta)=\sum_i \Delta F_{i}-\sum_a (k_a-1)\Delta F_a
\end{equation}
\begin{equation}
\begin{split}
\textrm{with} \quad \Delta F_i=\Delta F_{\circ+\square\in\circ}(\{u_{a\vers i}\}) &\equiv\frac{1}{\beta}\sum_{a\in i}\ln [2\cosh(\beta u_{a\to i})] -\frac{1}{\beta}\ln \left[2\cosh\left(\beta h_i+\beta\sum_{a\in i}u_{a\to i}\right)\right]\\
\Delta F_a=\Delta F_{\square}(\{h_{i\vers a}\}) &\equiv -\frac{1}{\beta}\ln\left(\frac{1+\prod_{i\in a}\tanh\beta h_{i\to a}}{2}\right)
\end{split}
\end{equation}

Define:
\begin{equation}
P(h)=\frac{1}{N\langle \ell\rangle}\E_C\left[\sum_{(i,a)}\delta(h-h_{i\vers a})\right]\qquad Q(u)=\frac{1}{N\langle \ell\rangle}\E_C\left[\sum_{(i,a)}\delta(u-u_{a\vers i})\right]
\end{equation}
Averaging \eqref{bpbsc} over the codes, the noise and the edges, we obtain the self-consistency equations:
\begin{eqnarray}
P(h)&=&\sum_\ell \frac{\ell v_\ell}{\langle \ell\rangle}\int \prod_{a=1}^{\ell-1}\ud u_a\,Q(u_a)\left\langle\delta\left[h-\hat h(h_\xi,\{u_a\})\right]\right\rangle_{h_\xi}\\
Q(u)&=&\sum_k \frac{k c_k}{\langle k\rangle}\int \prod_{i=1}^{k-1}P(h_i)\delta\left[u-\hat u(\{h_i\})\right]
\end{eqnarray}
where $h_\xi=h_0$ with probability $1-p$ and $-h_0$ with probability $p$. The RS free energy reads:
\begin{equation}
f_{RS}(\beta)=\sum_\ell v_\ell  \int \prod_{a=1}^\ell \ud u_a\,Q(u_a)\left\langle\Delta F_{\circ+\square\in\circ}(h_\xi,\{u_a\})\right\rangle_{h_\xi}-\sum_k c_k (k-1) \int \prod_{i=1}^k \ud h_i\,P(h_i)\Delta F_{\square}(\{h_i\})
\end{equation}

\subsection{Large Deviations}
As in the BEC, we study the statistics of BP over the codes, under the measure $\propto \exp[-x_f \beta_f F_\textrm{corr}(\beta_f)-x_e \beta_e F_\textrm{RS}(\beta_e)]$. The large deviation cavity equations read, for a regular code:
\begin{equation}\label{ldc}
\begin{split}
P(h)\propto&\int\prod_{a=1}^{\ell-1}\ud u_a\ Q(u_a) \frac{\left\langle \delta\left(h-h_\xi-\sum_{a=1}^{\ell-1} u_a\right)\e^{\beta_fx_fh_\xi}{\left[2\cosh\left(\beta_e (h_\xi+\sum_{a=1}^{\ell-1}u_a)\right)\right]}^{x_e}\right\rangle_{h_\xi}}{\prod_{a=1}^{\ell-1}{\left[2\cosh(\beta_e u_a)\right]}^{x_e}},\\
Q(u)=&\int\prod_{i=1}^{k-1}\ud h_i\ P(h_i)\delta\left[u-\frac{1}{\beta}\mathrm{atanh}\left(\prod_{i=1}^{k-1}\tanh(\beta_p h_i)\right)\right]
\end{split}
\end{equation}
And the potential:
\begin{equation}
\begin{split}
\phi(\beta_f,\beta_e,x_f,x_e)&=\ln\int\prod_{a=1}^\ell \ud u_a\ Q(u_a)\frac{\left\langle \e^{\beta_fx_fh_\xi}{\left[2\cosh\left(\beta_e (h_\xi+\sum_{a=1}^{\ell} u_a)\right)\right]}^{x_e}\right\rangle_{h_\xi}}{\prod_{a=1}^{\ell}{\left[2\cosh(\beta_e u_a)\right]}^{x_e}}\\
&-\frac{\ell}{k}(k-1)\ln \int\prod_{i=1}^{k}dh_i P(h_i){\left[\frac{1+\prod_{i=1}^k\tanh(\beta_e h_i)}{2}\right]}^{x_e}
\end{split}
\end{equation}

The solution to \eqref{ldc} is obtained numerically. In the limit $k,\ell\vers\infty$, this solution simplifies:
\begin{equation}\label{eq:typeII}
Q(u)=\delta(u)\qquad
P(h)=(1-p)\delta(h-h_0)+p\delta(h+h_0)
\end{equation}
yielding the error exponent \eqref{klinfinylimit}.

Another solution, called ``type I'' in \cite{SkantzosvanMourik03}, also exists:
\begin{equation}\label{eq:typeI}
Q(u)=\eta\delta_{+\infty}(u)+(1-\eta)\delta_{-\infty}(u)\qquad
P(h)=\nu\delta_{+\infty}(h)+(1-\nu)\delta_{-\infty}(h)
\end{equation}
with
\begin{equation}
\nu=\frac{\eta^{\ell-1}}{\eta^{\ell-1}+(1-\eta)^{\ell-1}\left\langle\e^{-2yh_0\sigma}\right\rangle_\sigma},\qquad
\eta=\frac{1}{2}\left(1+(2\nu-1)^{k-1}\right),
\end{equation}
We automatically have $s_p=0$, and the condition $f_p=f_f$ implies $m=\beta_e x_e=1/2$. Then the rate function reads:
\begin{equation}
L_1(f_p=f_f)=-\phi=-\ln \left[\eta^\ell+(1-\eta)^\ell\left\langle\e^{-h_0\sigma}\right\rangle_\sigma \right]-\frac{\ell}{k}(k-1)\ln\left[\frac{1}{2}(1+(2\nu-1)^k)\right]
\end{equation}
This solution \eqref{eq:typeI} is numerically unstable and the rate function thus obtained is clearly unphysical. However, for $k,\ell\to\infty$, $\ell/k=1-R$, we have $\eta=\nu=1/2$ and the resulting rate function
\begin{equation}
L_1(f_p=f_f)=-\ln\frac{1}{2}\left(1+2\sqrt{p(1-p)}\right)-R\ln 2=\ln 2(R_0(p)-R)
\end{equation}
coincides with the error exponent of the RLM in the low $p$ regime \eqref{RLMBSC}.

\subsection{Two-step large deviations}

The potential $\psi(\beta_e,m,y)$ defined in \eqref{eq:psibsc} is obtained by extremizing the following expression with respect to $P(h)$ and $Q(u)$:
\begin{equation}
\begin{split}
\psi(\beta_e,m,y)&=\ln\int\prod_{a=1}^\ell \ud u_a\ Q(u_a){\left\{\frac{\left\langle \e^{-mh_\xi}{\left[2\cosh\left(\beta_e (h_\xi+\sum_{a=1}^{\ell} u_a)\right)\right]}^{m/\beta_e}\right\rangle_{h_\xi}}{\prod_{a=1}^{\ell}{\left[2\cosh(\beta_e u_a)\right]}^{m/\beta_e}}\right\}}^y\\
&-\frac{\ell}{k}(k-1)\ln \int\prod_{i=1}^{k}dh_i P(h_i){\left[\frac{1+\prod_{i=1}^k\tanh(\beta_e h_i)}{2}\right]}^{ym/\beta_e}
\end{split}
\end{equation}

We can only handle this calculation in the $k,\ell\vers\infty$ limit. \eqref{eq:typeII} is still a solution in this case, and yields:
\begin{equation}
\psi(\beta_e,m,y)=y\hat\phi(\beta_e,m),
\end{equation}
where $\hat\phi(\beta_e,m)$ is obtained from the average case.
Therefore, the typical exponent is the same as the average error exponent in the high $p$ regime. 

There also exists a counterpart of solution \eqref{eq:typeI}, which gives:
\begin{equation}
\psi(\beta_e,m,y)=(R-1)\ln 2+\ln\left[1+\left((1-p)^{1-m}p^m+p^{1-m}(1-p)^m\right)^y\right]
\end{equation}
The condition $\partial_m \psi=0$ is again enforced by setting $m=1/2$. Thus we get:
\begin{equation}
\psi(y)=-yL-\L=(R-1)\ln 2+\ln\left[1+\left(2\sqrt{p(1-p)}\right)^y\right]
\end{equation}
This expression yields the rate function $\L(L)$ by inverse Legendre transformation.

\bibliographystyle{unsrt}
\bibliography{myentries}

\begin{thebibliography}{10}

\bibitem{Shannon48}
C.~E. Shannon.
\newblock A mathematical theory of communication.
\newblock {\em Bell System Tech. Journal}, 27:379--423, 623--655, 1948.

\bibitem{BerrouGlavieux93}
C.~Berrou, A.~Glavieux, and P.~Thitimajshima.
\newblock Near shannon limit error-correcting coding: Turbo codes.
\newblock In {\em Proc. IEEE International Conference on Communications}, pages
  1064--1070, 1993.

\bibitem{MacKay03}
D.~J.~C. MacKay.
\newblock {\em Information theory, inference, and learning algorithms}.
\newblock Cambridge University Press, Cambridge, 2003.

\bibitem{Gallager62}
R.~G. Gallager.
\newblock Low-density parity check codes.
\newblock {\em IRE Trans. Inf. Theory}, IT-8:21, 1962.

\bibitem{Verdu98}
S.~Verd{\'u}.
\newblock Fifty years of shannon theory.
\newblock {\em IEEE Transactions on Information Theory}, 44(6):2057--2078,
  1998.

\bibitem{Berlekamp02}
E.~R. Berlekamp.
\newblock The performance of block codes.
\newblock {\em Notices of the AMS}, pages 17--22, January 2002.

\bibitem{BargForney02}
A.~Barg and G.~D.~Forney Jr.
\newblock Random codes : minimum distances and error exponents.
\newblock {\em IEEE Trans. Inform. Theory}, 48:2568--2573, 2002.

\bibitem{DiProietti02}
C.~Di, D.~Proietti, I.~E. Telatar, and R.~L.~Urbanke T.~J.~Richardson.
\newblock Finite length analysis of low-density parity-check codes on the
  binary erasure channel.
\newblock {\em IEEE Trans. Inform. Theory}, 48:1570--1579, 2002.

\bibitem{AmraouiMontanari04}
A.~Amraoui, A.~Montanari, T.~Richardson, and R.~Urbanke.
\newblock Finite-length scaling for iteratively decoded ldpc ensembles.
\newblock Submitted IEEE Trans. on Information Theory. Preprint cs.IT/0406050,
  2004.

\bibitem{SkantzosvanMourik03}
N.~S. Skantzos, J.~van Mourik, D.~Saad, and Y.~Kabashima.
\newblock Average and reliability error exponents in low-density parity-check
  codes.
\newblock {\em J. Phys. A}, 36:11131--11141, 2003.

\bibitem{NishimoriBook01}
H.~Nishimori.
\newblock {\em Statistical Physics of Spin Glasses and Information Processing:
  An Introduction}.
\newblock Oxford University Press, Oxford, UK, 2001.

\bibitem{Sourlas89}
N.~Sourlas.
\newblock Spin-glass models as error-correcting codes.
\newblock {\em Nature}, 339:693--694, 1989.

\bibitem{denHollander00}
F.~den Hollander.
\newblock {\em Large deviations}.
\newblock Fields Institute Monographs 14. American Mathematical Society,
  Providence RI, 2000.

\bibitem{MezardParisi01}
M.~M{\'e}zard and G.~Parisi.
\newblock The bethe lattice spin glass revisited.
\newblock {\em Eur. Phys. J. B}, 20:217, 2001.

\bibitem{Rivoire05}
O.~Rivoire.
\newblock The cavity method for large deviations.
\newblock {\em J. Stat. Mech.}, page P07004, 2005.

\bibitem{MoraRivoire06a}
T.~Mora and O.~Rivoire.
\newblock Error exponents of low-density parity-check codes on the binary
  erasure channel.
\newblock cs.IT/0605130, 2006.

\bibitem{CoverThomas91}
T.~M. Cover and J.~A. Thomas.
\newblock {\em Elements of information theory}.
\newblock Wiley, New-York, 1991.

\bibitem{Tanner81}
R.~M. Tanner.
\newblock A recursive approach to low complexity codes.
\newblock {\em IEEE Trans. on Information Theory}, 27:533--547, 1981.

\bibitem{Bollobas01}
B.~Bollob{\'a}s.
\newblock {\em Random graphs}.
\newblock Cambridge University Press, second edition, 2001.

\bibitem{MezardParisi87b}
M.~M{\'e}zard, G.~Parisi, and M.~A. Virasoro.
\newblock {\em Spin-Glass Theory and Beyond}, volume~9 of {\em Lecture Notes in
  Physics}.
\newblock World Scientific, Singapore, 1987.

\bibitem{PapadimitriouSteiglitz82}
C.~H. Papadimitriou and K.~Steiglitz.
\newblock {\em Combinatorial Optimization: Algorithms and Complexity}.
\newblock Prentice Hall, Englewood Cliffs, NJ, 1982.

\bibitem{RicciWeigt01}
F.~Ricci-Tersenghi, M.~Weigt, and R.~Zecchina.
\newblock Simplest random $k$-satifiablitity problem.
\newblock {\em Phys. Rev. E}, 63:026702, 2001.

\bibitem{CoccoDubois03}
S.~Cocco, O.~Dubois, J.~Mandler, and R.~Monasson.
\newblock Rigorous decimation-based construction of ground pure states for spin
  glass models on random lattices.
\newblock {\em Phys. Rev. Lett.}, 90:047205, 2003.

\bibitem{MezardRicci03}
M.~M{\'e}zard, F.~Ricci-Tersenghi, and R.~Zecchina.
\newblock Alternative solutions to diluted $p$-spin models and {XORSAT}
  problems.
\newblock {\em J. Stat. Phys.}, 111:505, 2003.

\bibitem{MezardParisi03}
M.~M{\'e}zard and G.~Parisi.
\newblock The cavity method at zero temperature.
\newblock {\em J. Stat. Phys.}, 111:1--34, 2003.

\bibitem{KabashimaSaad04}
Y.~Kabashima and D.~Saad.
\newblock Statistical mechanics of low-density parity-check codes.
\newblock {\em J. Phys. A: Math. Gen}, 37:R1--R43, 2004.

\bibitem{Montanari01}
A.~Montanari.
\newblock The glassy phase of {G}allager codes.
\newblock {\em Eur. Phys. J. B.}, 23:121--136, 2001.

\bibitem{FranzLeone02}
S.~Franz, M.~Leone, A.~Montanari, and F.~Ricci-Tersenghi.
\newblock The dynamic phase transition for decoding algorithms.
\newblock {\em Phys. Rev. E}, 66:046120, 2002.

\bibitem{MezardPalassini05}
M.~M{\'e}zard, M.~Palassini, and O.~Rivoire.
\newblock Landscape of solutions in constraint satisfaction problems.
\newblock {\em Phys. Rev. Lett.}, 95:200202, 2005.

\bibitem{MontanariRicci03}
A.~Montanari and F.~Ricci-Tersenghi.
\newblock On the nature of the low-temperature phase in discontinuous
  mean-field spin glasses.
\newblock {\em Eur. Phys. J. B}, 33:339, 2003.

\bibitem{footnoteContrary}
Contrary to what indicates the last equations of~\cite{Rivoire05}, the nature
  of the order parameter is unchanged when additional levels of disorder are
  taken into account. The reason is that the cavity method encodes in a unique
  spatial distribution both the statistics over the nodes of a single graph,
  and the statistics over the graphs in a ensemble. The discrimination between
  the two levels is done only through the unequal weighting attributed to the
  different nodes, as controlled by the two independent temperatures $x$ and
  $y$.

\bibitem{Nattermann98}
T.~Nattermann.
\newblock Theory of the random field ising model.
\newblock In A.~P. Young, editor, {\em Spin Glasses and Random Fields}. World
  Scientific, Singapore, 1998.

\bibitem{MontanariSemerjian05}
A.~Montanari and G.~Semerjian.
\newblock From large scale rearrangements to mode coupling phenomenology.
\newblock {\em Phys. Rev. Lett.}, 94:247201, 2005.

\bibitem{MontanariSemerjian05-2}
A.~Montanari and G.~Semerjian.
\newblock On the dynamics of the glass transition on bethe lattices.
\newblock cond-mat/0509366.

\bibitem{MartinMezard05}
O.~C. Martin, M.~M{\'e}zard, and O.~Rivoire.
\newblock Random multi-index matching problems.
\newblock {\em J. Stat. Mech.}, page P09006, 2005.

\bibitem{MeassonMontanari04}
C.~Measson, A.~Montanari, T.~Richardson, and R.~Urbanke.
\newblock Life above threshold: from list decoding to area thereom and {MSE}.
\newblock In {\em Proc. ITW}, San Antonio, USA, October 2004.

\bibitem{RivoireBarre06}
O.~Rivoire and J.~Barr{\'e}.
\newblock Exactly solvable models of adaptive networks.
\newblock In preparation, 2006.

\bibitem{MourikKabashima03}
J.~van Mourik and Y.~Kabashima.
\newblock The polynomial error probability for ldpc codes.
\newblock cond-mat/0310177, 2003.

\bibitem{Gallager68}
R.~G. Gallager.
\newblock {\em Information theory and reliable communication}.
\newblock John Wiley and Sons, New York, 1968.

\bibitem{Condamin02}
S.~Condamin.
\newblock Study of the weight enumerator function for a gallager code.
\newblock http://www.inference.phy.cam.ac.uk/condamin/report.ps, 2002.

\bibitem{DiMontanari04}
C.~Di, A.~Montanari, and R.~Urbanke.
\newblock Weight distributions of {LDPC} code ensembles: Combinatorics meets
  statistical physics.
\newblock In {\em International Symposium on Information Theory}. IEEE, 2004.

\bibitem{footnoteInourcase}
In our case
  $v^{(k)}_\ell=\sum_{\ell'\geq\ell}v_{\ell'}\binom{\ell'}{\ell}c_k^\ell(1-c_k%
)^{\ell'-\ell}$.

\end{thebibliography}

\end{document}